\documentclass[aps, pra, twocolumn, superscriptaddress]{revtex4-2}

\usepackage{hyperref, graphicx, xcolor, amsmath, amssymb, siunitx}
\usepackage[caption=false]{subfig}
\usepackage{verbatim}
\usepackage{braket}
\hypersetup{colorlinks, linkcolor=teal, citecolor=blue, urlcolor=purple}

\begin{document}

\title{One hundred second bit-flip time in a two-photon dissipative oscillator}
\author{C.~Berdou}
\affiliation{Laboratoire de Physique de l’Ecole normale supérieure, Mines-Paristech, Inria, ENS-PSL, Université PSL, CNRS, Sorbonne Université, Paris, France}
\author{A.~Murani}
\affiliation{Alice \& Bob, 53 Bd du Général Martial Valin, 75015 Paris, France}
\author{U.~Réglade}
\affiliation{Alice \& Bob, 53 Bd du Général Martial Valin, 75015 Paris, France}
\affiliation{Laboratoire de Physique de l’Ecole normale supérieure, Mines-Paristech, Inria, ENS-PSL, Université PSL, CNRS, Sorbonne Université, Paris, France}
\author{W.~C.~Smith}
\author{M.~Villiers}
\author{J.\ Palomo}
\author{M.\ Rosticher}
\author{A.\ Denis}
\author{P.\ Morfin}
\author{M.\ Delbecq}
\author{T.\ Kontos}
\affiliation{Laboratoire de Physique de l’Ecole normale supérieure, Mines-Paristech, Inria, ENS-PSL, Université PSL, CNRS, Sorbonne Université, Paris, France}
\author{N.\ Pankratova}
\author{F.\ Rautschke}
\author{T.\ Peronnin}
\affiliation{Alice \& Bob, 53 Bd du Général Martial Valin, 75015 Paris, France}
\author{L.-A.~Sellem}
\author{P.~Rouchon}
\author{A.~Sarlette}
\author{M.~Mirrahimi}
\author{P.~Campagne-Ibarcq}
\affiliation{Laboratoire de Physique de l’Ecole normale supérieure, Mines-Paristech, Inria, ENS-PSL, Université PSL, CNRS, Sorbonne Université, Paris, France}
\author{S.~Jezouin}
\author{R.~Lescanne}
\affiliation{Alice \& Bob, 53 Bd du Général Martial Valin, 75015 Paris, France}
\author{Z.~Leghtas}
\email[]{zaki.leghtas@ens.fr}
\affiliation{Laboratoire de Physique de l’Ecole normale supérieure, Mines-Paristech, Inria, ENS-PSL, Université PSL, CNRS, Sorbonne Université, Paris, France}
\date{\today}


\begin{abstract}
Current implementations of quantum bits (qubits) continue to undergo too many errors to be scaled into useful quantum machines. An emerging strategy is to encode quantum information in the two meta-stable pointer states of an oscillator  exchanging pairs of photons with its environment, a mechanism shown to provide stability without inducing decoherence. Adding photons in these states increases their separation, and macroscopic bit-flip times are expected even for a handful of photons, a range suitable to implement a qubit. However, previous experimental realizations have saturated in the millisecond range. In this work, we aim for the maximum bit-flip time we could achieve in a two-photon dissipative oscillator. To this end, we design a Josephson circuit in a regime that circumvents all suspected dynamical instabilities, and employ a minimally invasive fluorescence detection tool, at the cost of a two-photon exchange rate dominated by single-photon loss. We attain bit-flip times of the order of 100 seconds for states pinned by two-photon dissipation and containing about 40 photons. This experiment lays a solid foundation from which the two-photon exchange rate can be gradually increased, thus gaining access to the preparation and measurement of quantum superposition states, and pursuing the route towards a logical qubit with built-in bit-flip protection.
\end{abstract}
\maketitle

\section{Introduction}
\begin{figure}
\centering
\includegraphics[width=1\columnwidth]{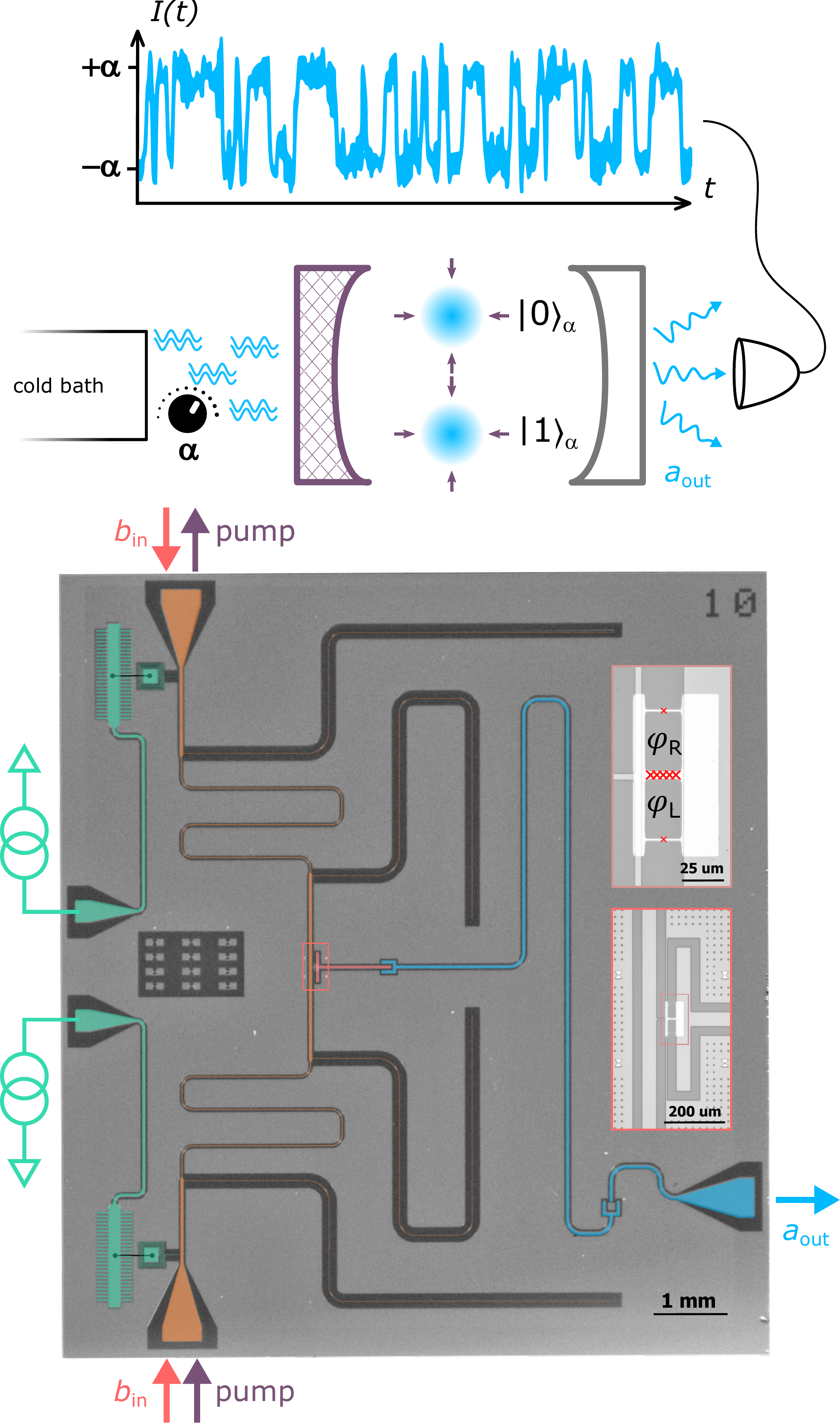}
\caption{(Top) Principle of the experiment. A cavity is endowed with a special mechanism (dashed left mirror) that exchanges pairs of photons (blue double waves) at variable intensity (control knob) with a cold bath. Two meta-stable pointer states emerge, represented by the blue distributions centered around amplitudes $\pm\alpha$. A fraction of the cavity field (blue waves) escapes through the weakly transmissive mirror and is collected by our heterodyne detector. By monitoring the signal over time, we track individual trajectories undergoing bit-flips (blue time trace). (Bottom) False-color optical micrograph of the experimental superconducting circuit in a coplanar waveguide geometry. The cavity is a $\lambda/2$ resonator (blue) that radiates a field $a_\text{out}$ through a weakly coupled $50~\Omega$ port. It also couples to a two-photon exchange device composed of a buffer mode (red) shunted to ground through an asymmetrically threaded SQUID (ATS) as emphasized in the insets. {DC currents enter through on-chip bias tees (green) and impose phase biases $\varphi_{L,R}$. A differential pump (purple arrows) and a common buffer drive (red arrows) are channeled through filtered transmission lines (orange).}}
\label{fig:Fig1}
\end{figure}

The quest for a physical system suitable for quantum information processing is intensifying at the dawn of quantum computing. Microscopic entities such as single ions \cite{Bruzewicz2019} and electrons \cite{Vandersypen2019} came forward as initial promising candidates. More recently, human-made superconducting circuits encode quantum bits in the form of single excitations of electromagnetic modes \cite{Blais2021}. Despite impressive progress \cite{Arute2019}, these systems continue to be individually plagued by too many errors, blocking their deployment into scalable quantum machines.


A qubit with computational states $\ket{0}$ and $\ket{1}$ undergoes errors spanned by two processes: bit-flips that randomly swap $\ket{0}$ and $\ket{1}$, and phase-flips that scramble the phase of quantum superpositions of $\ket{0}$ and $\ket{1}$. Unlike phase-flips, bit-flips have a clear classical analogue. Interestingly, classical bits in a typical static random access memory have bit-flip times in the range of $10^{15}$ seconds \cite{Autran2014}, 17 orders of magnitude larger than their quantum counterpart \cite{Earnest2018}. This observation sparked interest in a qubit whose computational states would be stable over macroscopic time-scales.

A nonlinear dynamical system that is open to its environment through a carefully tailored interaction may exhibit a rich dynamical phase space hosting multiple stable steady states \cite{Dykman1980, Guckenheimer1983}. The switching between these states in dissipative Kerr oscillators has long been used for amplification \cite{siddiqi2004, Krantz2016} and ultra-low power classical logic \cite{Mabuchi2011, Kerckhoff2012}. Their stability is ensured by regular energy damping, that, at the same time, decoheres their quantum superpositions. Surprisingly, there exists a mechanism, known as two-photon dissipation, that provides stability without inducing decoherence. A recent qubit, coined the cat-qubit \cite{Mirrahimi2014}, is embedded in the cavity field of a superconducting resonator that exchanges pairs of photons with its environment \cite{Carmichael2008, Leghtas2015}. This defies the common intuition that a qubit must be well isolated from its environment. Instead, the subtle interplay of drive and dissipation pins down the cavity field on two coherent steady states with complex amplitudes denoted $\pm\alpha$ without affecting quantum superpositions of the two.


Increasing the number of photons $\bar{n}=|\alpha|^2$ in these two steady states has two opposing effects \cite{Lescanne2020}. On the one hand, their distinguishability by an inevitably coupled uncontrolled environment increases linearly with $\bar{n}$. This results in a linear {increase of the phase-flip error rate}. Therefore, for this system to be suitable for quantum information processing, it must operate at low photon number. On the other hand, as soon as their separation exceeds their vacuum fluctuations, that is $|\alpha-(-\alpha)|^2=4{\bar{n}}>1$, their wave-function overlap rapidly decreases, reducing random tunneling between them and hence exponentially increasing the bit-flip time. It is remarkable that, at least in principle, it is possible to reach macroscopic bit-flip times with computational states pinned by two-photon dissipation containing only a handful of photonic excitations.

Previous experiments have succeeded in implementing a two-photon exchange mechanism to observe the squeezing of a Schr\"odinger cat state out of vacuum \cite{Leghtas2015}, the dynamics of a quantum gate \cite{Touzard2018}, the exponential suppression of bit-flips and linear increase of phase-flips \cite{Lescanne2020}. A recent insight has been to define a cat-qubit out of the interplay of Kerr non-linearity and single mode squeezing \cite{Puri2017,Grimm2020}. However, in all these implementations, the bit-flip time saturated in the millisecond range, limited by errors impinging from the cat-qubit tomography apparatus \cite{Lescanne2020}, and possible dynamical instabilities \cite{Lescanne2019, Verney2019}. 




In this work, we aim for the maximum bit-flip time we could achieve in a two-photon dissipative oscillator. To reach this goal, we first design a circuit with the objective of removing all suspected sources of dynamical instabilities and ancillary systems that could propagate uncorrectable errors. We fabricate a two-photon exchange dipole element close to the regime where its energy landscape exhibits a single global minimum at any operating point, a possible requirement for stability \cite{Verney2019, Burgelman2022}. Second, we entirely remove the tomography apparatus: our design does not contain a transmon and readout mode. Instead, we directly measure the field radiated by the cavity through a travelling wave parametric amplifier (TWPA) \cite{Macklin2015}, thereby accessing individual oscillator state trajectories. We measure a bit-flip time exceeding 100 seconds for computational states pinned by two-photon dissipation and containing about 40 photons. Our design choices came at the cost of a two-photon exchange rate dominated by single-photon loss, hence losing our ability to prepare quantum superposition states and hence measuring the phase-flip rate. Guided by this benchmark, future experiments can then gradually enter the regime suitable to implement a qubit where two-photon loss is the dominant dissipation mechanism.

\section{The two-photon dissipative oscillator}
An oscillator exchanging pairs of photons with its environment in addition to usual energy relaxation (see Fig.~\ref{fig:Fig1}) is modeled by the following Hamiltonian and loss operators:
\begin{equation}
H_2/\hbar={i}\epsilon_2a^{\dag2}-{i}\epsilon_2^*a^2,\;\;L_2 =\sqrt{\kappa_2}a^2\;,\;\; L_1 = \sqrt{\kappa_a}a\;,
\label{eq:L2L1}
\end{equation}
where $a$ is the annihilation operator of the mode referred to as the memory, $\epsilon_2$ is the two-photon injection rate, $\kappa_2$ the two-photon loss rate, and $\kappa_a$ is the energy damping rate. When the two-photon injection rate overcomes the damping rate, two meta-stable pointer states emerge:
\begin{equation*}
\ket{0}_\alpha = \ket{+\alpha} + \mathcal{O}(e^{-2|\alpha|^2})\;,\;\;
\ket{1}_\alpha = \ket{-\alpha} + \mathcal{O}(e^{-2|\alpha|^2})\;,
\label{eq:01}
\end{equation*}
where $\ket{\alpha}$ is a coherent state with complex amplitude $\alpha$, verifying
$
    \alpha^2 = \frac{2}{\kappa_2}(\epsilon_2-\kappa_a/4) \text{ if } \epsilon_2>\kappa_a/4, \text{ and } \alpha^2=0 \text{ otherwise}.
$

The two-photon dissipation mechanism is engineered by implementing a two-to-one photon exchange interaction with a dissipative mode referred to as the buffer \cite{Leghtas2015}, modeled by the Hamiltonian
\begin{equation}
\label{eq:Hab}
    {H}_{ab}/\hbar = g_2^*a^2{b}^\dag + g_2a^{\dag 2}{b}-\epsilon_d b^\dag-\epsilon_d^* b\;,
\end{equation}
where ${b}$ is the annihilation operator of the buffer, $g_2$ is the two-to-one photon coupling rate and $\epsilon_d$ the buffer drive amplitude. In the limit where the buffer energy decay rate $\kappa_b$ is larger than $|g_2|$, we recover Eq.~\eqref{eq:L2L1} with
$\kappa_2={4|g_2|^2}/{\kappa_b}$ and $\epsilon_2 = 2g_2\epsilon_d/\kappa_b$ \cite{Leghtas2015}.
\begin{figure*}
    \centering
    \includegraphics[width=1.95\columnwidth]{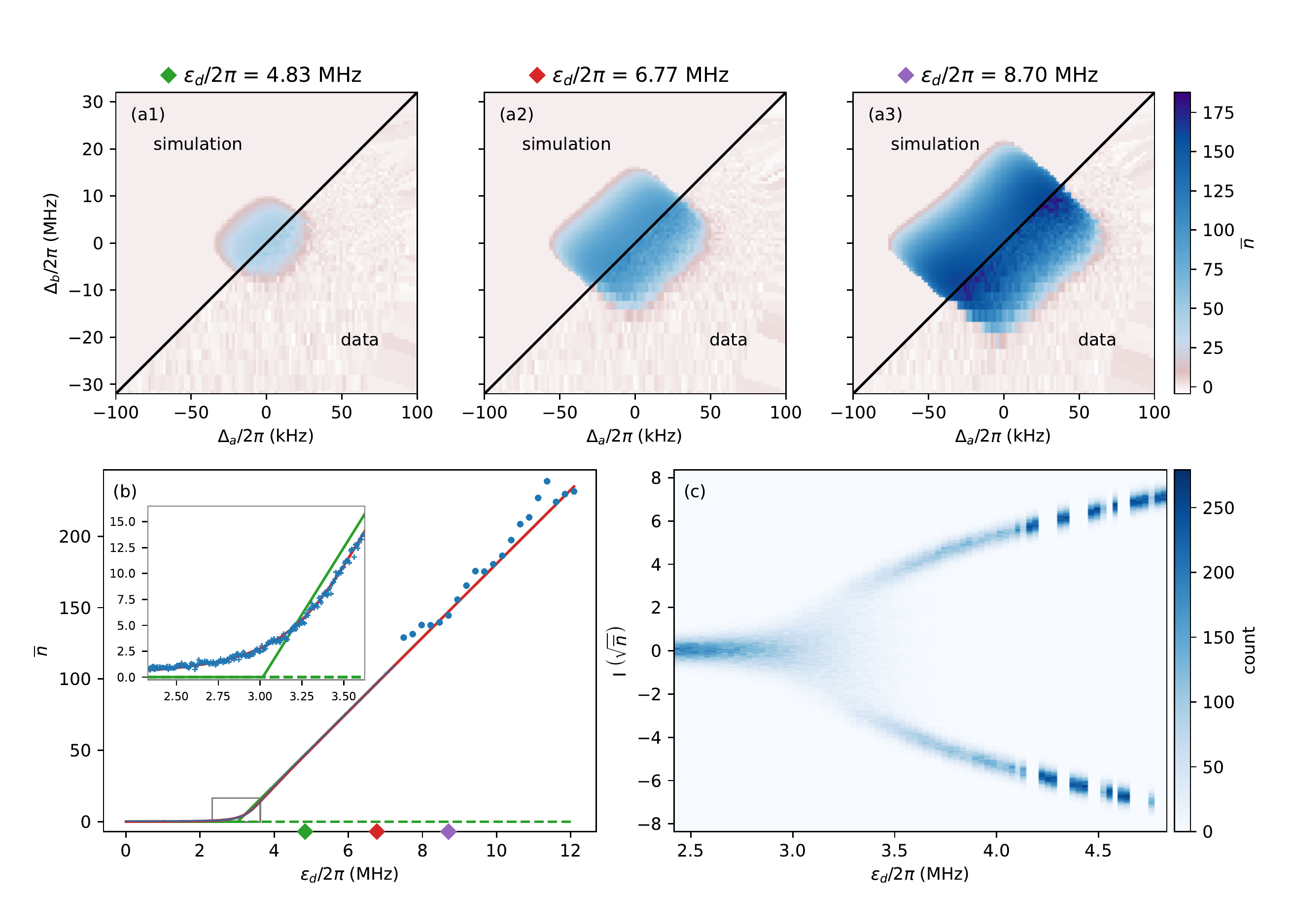}
    \caption{Emergence of two meta-stable pointer states from a nonlinear dissipative phase transition. (a1-a3) Radiated energy from the memory mode in units of circulating photon number (color) as a function of the detuning from the frequency matching condition $\Delta_a=\frac{1}{2}\left(\omega_p-(2\omega_a-\omega_d)\right)$ (x-axis), and the detuning from the buffer resonance $\Delta_b=\omega_d-\omega_b$ (y-axis). We denote $\omega_{a,b,p,d}$ the memory, buffer, pump and drive angular frequencies respectively. Both data and semi-classical numerical simulations are shown in different regions of each panel corresponding to the specified drive amplitude $\epsilon_d$. (b) Radiated energy from the memory in units of circulating photon number (y-axis) at $\Delta_a=\Delta_b=0$ as a function of the drive amplitude (x-axis). The data correspond to an integration time of $10~\mu s$ with single averaging (circles) and 10000 averages (crosses). A semi-classical model (green solid line) captures the appearance of a critical point around ${\epsilon_d / 2\pi\approx 3~\text{MHz}}$ above which the vacuum state becomes unstable (green dashed line). A full quantum model (red solid line) is necessary to capture the curvature at the critical point, as emphasized by the zoom in the inset panel. (c) Histogram (color) of the $I$-quadrature integrated over 1~ms of the field radiated by the memory (y-axis) in units of the square root of circulating photon number as a function of the drive amplitude (x-axis). Passed the critical point, the memory field transits from the vacuum into two meta-stable pointer states.
    }
    \label{fig:Fig2}
\end{figure*}

\section{Circuit design}
Our two-photon dissipative oscillator is implemented in a circuit quantum electrodynamics coplanar waveguide architecture (see Fig.\ref{fig:Fig1}). The memory mode is the fundamental mode of a $\lambda/2$ resonator. We measure {coupling and internal loss rates $\kappa_a^c/2\pi=40$~kHz and $\kappa_a^i/2\pi=18$~kHz}. In order to minimize dielectric losses \cite{Wang2015}, we target the relatively low frequency of $\omega_a/2\pi=4.0457$~GHz. A thermal population of about $1\%$ was measured on a twin sample using a transmon (see \cite{supmat} Sec.~\ref{sec:detection_efficiency}). An undesired side effect of coupling the memory to a lossy mode -- the buffer -- is to increase the decay rate of the memory due to the Purcell effect. To prevent this, we designed a stop-band filter centered at the memory frequency, consisting of three $\lambda/4$ sections \cite{Pozar2012} on both routes linking the memory to its cold bath. The buffer mode consists of a metallic plate of charging energy $E_C/h=73$~MHz shunted to ground through an Asymmetrically Threaded SQUID (ATS) \cite{Lescanne2019}. The ATS is formed by two Josephson junctions in a loop -- each of Josephson energy $E_J/h=37$~GHz -- split in its center by an inductance made of five junctions of total inductive energy $E_L/h=~62$~GHz. This layout defines two loops that can be biased in DC flux $\varphi_{L,R}$. We can hence independently control the common and differential flux through the ATS: $\varphi_\Sigma=\frac{1}{2}\left(\varphi_L+\varphi_R\right)$ and $\varphi_\Delta=\frac{1}{2}\left(\varphi_L-\varphi_R\right)$. Radio-frequency (RF) signals are routed to the ATS through a 180° hybrid coupler. The buffer drive propagates in phase through both arms of the two photon exchange apparatus. When reaching the ATS, these waves combine, inducing currents in the inductance and thereby driving the buffer mode. On the other hand, the pump propagates with opposite phases, inducing common flux in the ATS.


In the process of choosing the ATS parameters, we were guided by the intuition that dynamical instabilities would be avoided in a system with $2E_J/E_L \lesssim 1$ \cite{Verney2019, Burgelman2022}. However, this criterion needs to be balanced with the requirement of large $g_2$ (see \cite{supmat}, Sec.~\ref{section:supmat_analysis}). In this experiment, we favoured stability over coupling strength and chose: $2E_J/E_L=1.2$, a factor of $3.3$ smaller than our previous implementation \cite{Lescanne2019}. Moreover, we engineered a weak hybridization between the memory and buffer mode in order to minimize undesired nonlinear couplings such as the Kerr effect, with a rate estimated below $1$~Hz.

Previous experiments constructed the Wigner distribution of the memory field through a non-linear coupling to a transmon qubit and its readout resonator \cite{Lescanne2019}. However, the finite thermal occupation of the transmon was suspected to limit the bit-flip time to the millisecond range. Instead, we monitor our memory through a minimally invasive {detection} tool: a weakly coupled transmission line connected to a TWPA. This added leakage channel slightly decreases the total quality factor but has the advantage of not inducing any {additional} non-linear couplings to a lossy ancillary system.

\section{Experiment calibration}
The ATS contributes a non-linear potential energy of the form $U_{\varphi_\Sigma,\varphi_\Delta}(\varphi)=\frac{1}{2}E_L\varphi^2-2E_J\cos(\varphi_\Sigma)\cos(\varphi+\varphi_\Delta)$, where $\varphi$ is the phase drop across the central inductance \cite{Lescanne2019}. The buffer and memory modes hybridize through their capacitive coupling, and hence to the lowest order of their hybridization strength $\upsilon$, we have $\varphi=\varphi_b(b+b^\dag+\upsilon (a+a^\dag))$, where the buffer zero-point phase fluctuations verifies $\varphi_b=\left(2E_C/E_L\right)^{1/4}$ \cite{Girvin2014}. We operate the ATS at $\varphi_\Sigma=-\pi/2+\varepsilon_p\cos(\omega_p t)$, and $\varphi_\Delta=\pi/2$. At this operating point the buffer resonates at $\omega_b/2\pi=6.1273$~GHz with an energy decay rate $\kappa_b/2\pi=16$~MHz. By tuning the pump frequency at $\omega_p=2\omega_a-\omega_b$ and driving the buffer mode at $\omega_d=\omega_b$, we synthesize the Hamiltonian of Eq.~\eqref{eq:Hab}, with $\hbar g_2=-\frac{1}{2}E_J\varepsilon_p\upsilon^2\varphi_b^3$.

We start the experiment by measuring the buffer mode frequency map as a function of the two DC currents. Conveniently, the desired operating point $(\varphi_\Sigma, \varphi_\Delta)=(-\pi/2, \pi/2)$ is easily recognisable since it corresponds to a saddle point of this map (see \cite{supmat}, Sec.~\ref{section:supmat_analysis}). Consequently, the buffer and memory modes are first order insensitive to flux noise. 

The next step is to activate the RF pump {and buffer drive}. We pick the largest pump power that does not deteriorate the buffer and memory modes spectra. The modes frequencies are {Stark shifted} in the presence of this strong pump. Therefore, a precise calibration of the pump and drive frequencies is required to rigorously verify the frequency matching conditions: $\omega_p=2\omega_a-\omega_d$, and $\omega_d=\omega_b$. To this end, we acquire the memory mode fluorescence as a function of detunings from these matching conditions. As the drive amplitude $\epsilon_d$ is increased, the region over which the drive and pump combine to populate the memory expands around the frequency matching point (see Fig.~\ref{fig:Fig2}a). For the remaining of the experiment, we place ourselves at the center of these regions (colloquially referred to as diamonds).

We calibrate the number of photons in the memory by {measuring} $\overline{I^2+Q^2}$ (see \cite{supmat}, Sec.~\ref{section:supmat_photon_number_calibration}) as a function of the drive amplitude, {where $I$ and $Q$ are the in-phase and out-of-phase quadratures of the radiated field acquired over an integration time $T_m$.} (see Fig.~\ref{fig:Fig2}b). Three notable features are apparent. First, in the limit of strong drives, the radiated energy scales linearly with the drive amplitude, {a signature of the conversion of 1 buffer photon to 2 memory photons.} This is in stark contrast with the common quadratic scaling for a driven harmonic oscillator. Moreover, the offset of this asymptote from the origin excludes the Kerr effect as the underlying process (see \cite{supmat}, Sec.~\ref{section:steadystatephotonnumber}). Second, the output power is close to zero up until a critical drive amplitude, reminiscent of a non-linear dissipative phase transition \cite{Mylnikov2021}. This transition occurs when the two-photon injection rate overcomes the memory losses, setting the scale for the product $\epsilon_d g_2$ to its value at the critical point $\epsilon_d g_2=\kappa_a\kappa_b/8$.
Furthermore, we compute the classical dependence of $\bar{n}g_2^2$ on $\epsilon_d g_2$, demonstrating that it is invariant under a change of $g_2$ (see \cite{supmat}, Sec.~\ref{section:supmat_photon_number_calibration}). Importantly, quantum fluctuations blur the transition out of vacuum resulting in a non-scale invariant curvature, a striking deviation from the sharp transition expected in the classical regime. A full quantum model is necessary to capture this third notable feature, from which we extract the key parameter $g_2/2\pi=39$~kHz, and deduce $\kappa_2/2\pi=370$~Hz and $\bar{n}$ for every $\epsilon_d$. This places our experiment in the regime where $\kappa_a/\kappa_2=150\gg 1$. In the future we will increase the hybridization factor $\upsilon$ to enter the regime suitable for a qubit implementation: where the two-photon exchange rate largely dominates the cavity losses. Finally, with the photon number calibration in hand, the measurement records $I$ and $Q$ are rescaled to respectively coincide with a measurement of $(a+a^\dag)/2$ and $(a-a^\dag)/2i$.

\begin{figure*}
    \centering
    \includegraphics[width=2\columnwidth]{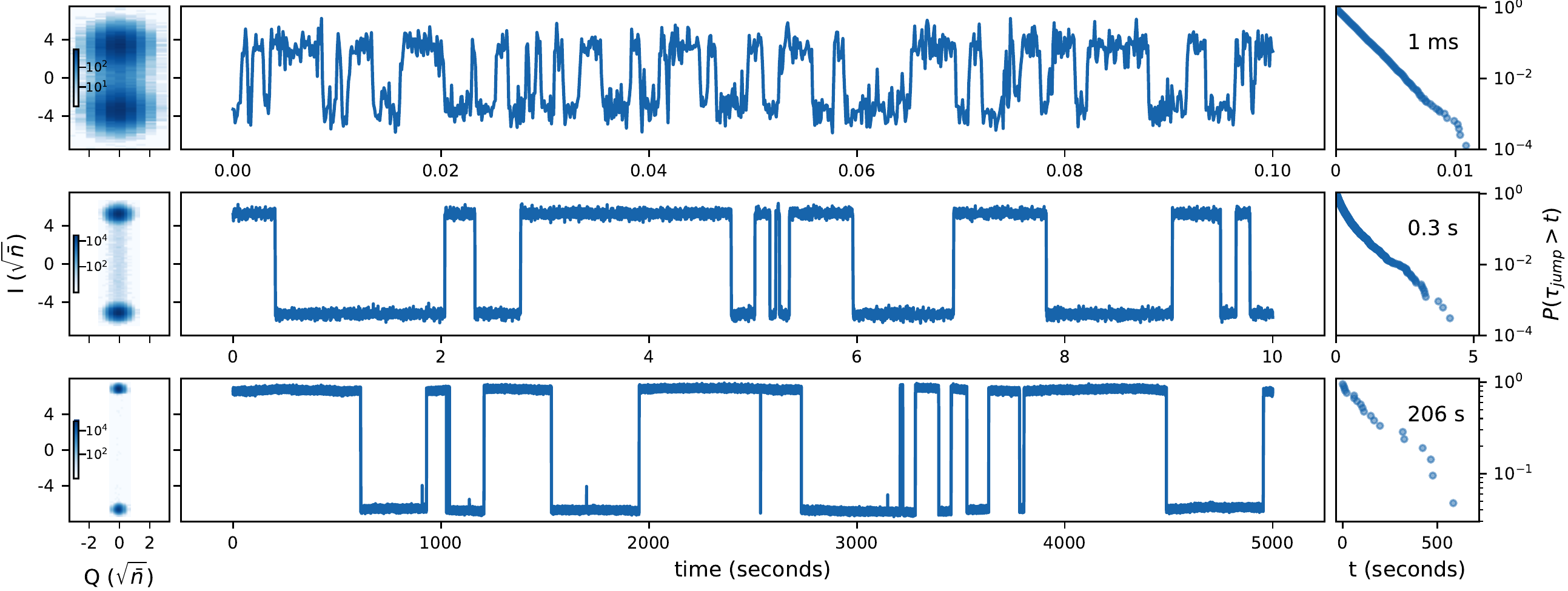}
    \caption{Real-time oscillator dynamics revealed by individual trajectories. For photon numbers $\bar{n}=11, 28, 43$ (top, middle, bottom), we respectively set the integration time to $T_m=0.1, 1, 5$~ms and the total measurement duration to $T_{tot}=10, 1000, 5000$~s. (Left) Histogram of the $(I,Q)$ quadratures of the radiated field. (Center) Trajectory of the $I$-quadrature as a function of time cropped from the full data-set. (Right) Cumulative distribution function of the stochastic time interval $\tau_\text{jump}$ in between two consecutive jumps. Its average value, that defines the bit-flip time, is printed on each panel.}
    \label{fig:Fig3}
\end{figure*}

In fact, the observed phase transition corresponds to a spontaneous symmetry breaking, where the cavity field adopts two opposite phases (or any quantum superposition of the two in the absence of single photon loss). We observe the emergence of these two phases by continuously acquiring, for each drive amplitude, 10000 times the $I$-quadrature of the radiated field integrated over $T_m=1$~ms, for a total measurement duration of 10~s (see Fig~\ref{fig:Fig2}c). For the lowest drive amplitudes ($\epsilon_d/2\pi\lesssim 2.7$~MHz), the cavity state remains in the vacuum, as signaled by the Gaussian distribution centered at $I=0$. This distribution then broadens around the critical point ($2.7~\text{MHz}\lesssim\epsilon_d/2\pi\lesssim 3.5~\text{MHz}$), due to the significant overlap of the distributions of states $\ket{\pm\alpha}$ at small $\alpha$ and the multiple flips in between during the acquisition time $T_m=1$~ms. For ($3.5~\text{MHz}\lesssim\epsilon_d/2\pi\lesssim 4~\text{MHz}$), the two states are well resolved, and their approximately equal weights hint towards a bit-flip time larger than the acquisition time of $1$~ms and smaller than the full experiment duration of 10~s. For $\epsilon_d/2\pi\gtrsim4$~MHz, the field stays pinned to one of the two computational states, hinting towards bit-flip times exceeding 10~s.


\section{Trajectories and bit-flip times}

We access the dynamics of the memory by tracking individual trajectories over time (see Fig.~\ref{fig:Fig3}). For each trajectory, we set the drive amplitude at a fixed value $\epsilon_d$, and record the $I$-quadrature of the radiated field. In order to resolve quantum jumps, we set the integration time $T_m$ to be simultaneously smaller than the bit-flip time  and sufficiently large to average out the heterodyne detection noise. To capture the statistical properties of each trajectory, we plot the cumulative distribution function of the interval between two consecutive jumps, denoted $\tau_\text{jump}$. It shows approximately an exponential law, revealing an underlying Poisson process. The average of $\tau_\text{jump}$, that defines the bit-flip time, undergoes a spectacular increase from 1~ms to 0.3~s to 206~s for an increase of photon number $\bar{n}$ from 11 to 28 to 43. With respect to the bare cavity lifetime of $2.7 {\rm \mu}$s, this represents a $10^8$ increase of the bit-flip time, and (although inaccessible with our measurement scheme) an estimated $2\times 43 = 86$ fold decrease of the phase-flip time. 

\begin{figure}
    \centering
    \includegraphics[width=0.9\columnwidth]{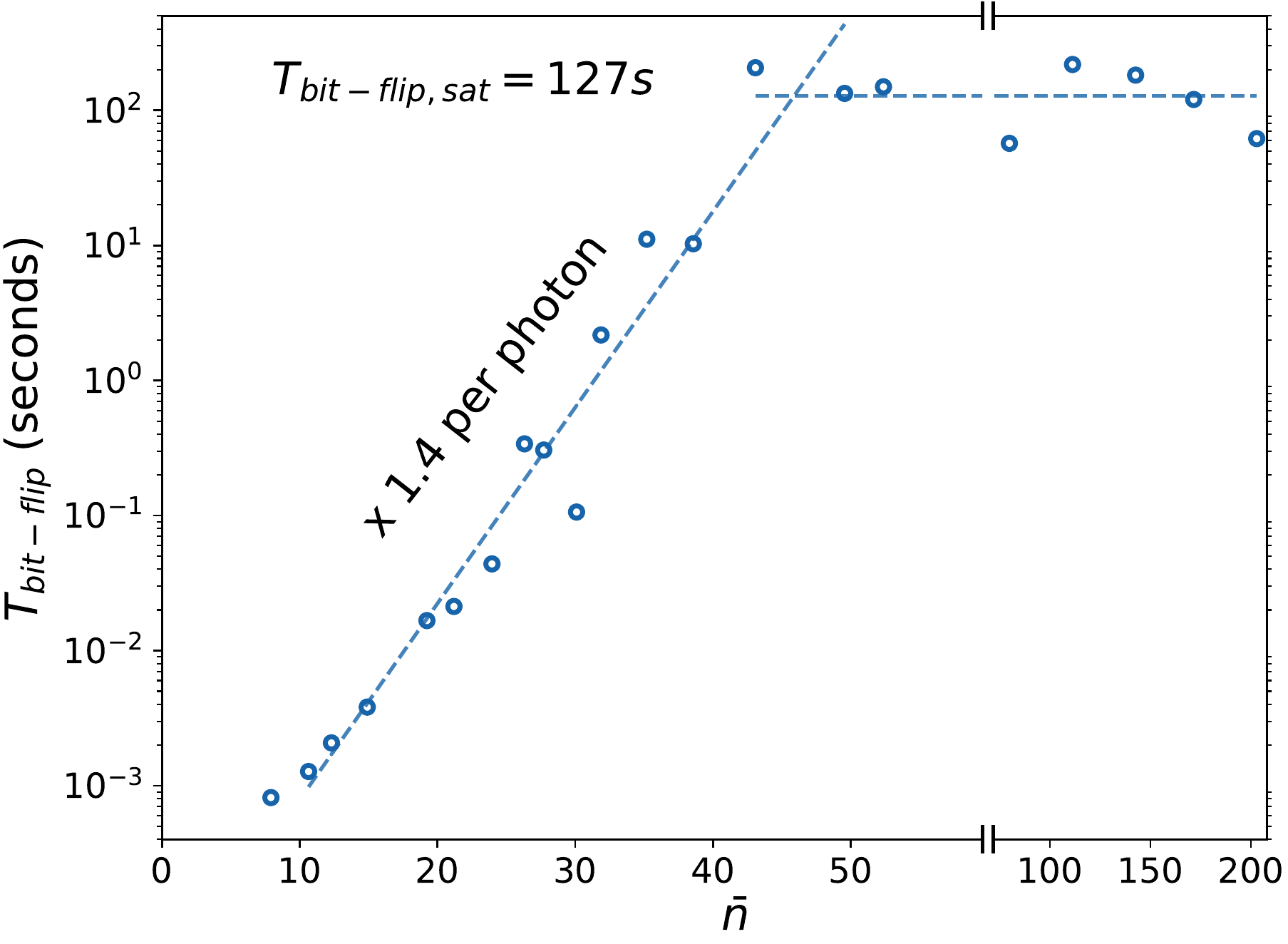}
    \caption{Exponential suppression of bit-flips. The bit-flip time (y-axis, log-scale) is measured (open circles) as a function of the number of photons contained in the pointer states $\ket{0/1}_\alpha$ (x-axis). The bit-flip time increases exponentially, multiplying by 1.4 per photon (tilted dashed line) before saturating at approximately 127 s (horizontal dashed line).}
    \label{fig:Fig4}
\end{figure}

We quantify the scaling of the bit-flip time with the photon number by repeating the trajectory acquisition procedure for multiple drive amplitudes. From each trajectory we extract the bit-flip time and the corresponding photon number, and display them in Fig.~\ref{fig:Fig4}. We observe two distinct regimes. Initially, the bit-flip time rises exponentially multiplying by a factor of about $1.4$ for every added photon. In theory, in the limit where $\kappa_a/\kappa_2\ll 1$, this factor would approach $e^{2}\sim 7.4$ \cite{Mirrahimi2014}. In this experiment we favoured stability over coupling strength, placing ourselves in the opposite regime $\kappa_a/\kappa_2\sim 150$, which is expected to decrease this factor as confirmed by numerical simulations (see \cite{supmat}, Sec.~\ref{section:supmat_bitflip_time}). For $\bar{n}\gtrsim 40$ photons, the bit-flip time saturates in the 100 second range. Although the origin of this saturation is yet to be established, its timescale is compatible with the measured rate of highly correlated errors in a large array of qubits \cite{McEwen2021}, possibly due to high energy particle impacts \cite{Vepslinen2020, Cardani2021}.


\section{Conclusion}

In conclusion, we have measured timescales of order 100 seconds for bit-flips between pointer states of a two-photon dissipative oscillator containing about 40 photons. To reach these macroscopic bit-flip times with mesoscopic photon numbers, we designed a two-photon exchange circuit in a regime expected to circumvent dynamical instabilities, and employed a minimally invasive detection tool that collects the oscillator's radiated field. {Our experiment thus puts a scale on the bit-flip times attainable with two-photon dissipation, a necessary mechanism for quantum information processing with cat-qubits \cite{Gautier2021}}. Future work could be to uncover the phenomena causing these bit-flip events \cite{Vool2014, Vepslinen2020, Cardani2021} by monitoring oscillator trajectories over timescales of days or weeks. Also, gradually increasing the two-photon exchange rate so that it supersedes all loss mechanisms would lead to the regime suitable for the cat-qubit where quantum superposition states can be prepared and measured, thereby paving the way towards fully protected chains of cat-qubits \cite{Guillaud2019, Chamberland2022}.


\section{Acknowledgments}
We thank Lincoln Labs for providing a Josephson Traveling-Wave Parametric Amplifier. The devices were fabricated within the consortium Salle Blanche Paris Centre. This work was supported by the QuantERA grant QuCOS, by ANR 19-QUAN-0006-04. Z.L.\ acknowledges support from ANR project ENDURANCE, and EMERGENCES grant ENDURANCE of Ville de Paris. This work has been supported by the Paris \^{I}le-de-France Region in the framework of DIM SIRTEQ. This project has received funding from the European Research Council (ERC) under the European Union’s Horizon 2020 research and innovation programme (grant agreements No.\ 851740 and No. 884762).


\section{Author contributions}
C.B simulated and fabricated the device. C.B, A.M and Z.L measured the device. C.B, A.M, U.R, R.L, S.J and Z.L analyzed the data and co-wrote the manuscript with input from all authors. W.C.S, M.V, A.D and P.M designed the sample holder. Support was provided by J.P, N.P and M.R for nanofabrication, M.D, T.K, R.L and T.P for experimental tools, F.R for microwave engineering, L.A.S, P.R, A.S and M.M for theory. Z.L, R.L, P.C and M.M conceived the experiment.

\bibliography{biblio}

\clearpage
\onecolumngrid
\begin{center}
\textbf{\Large Supplementary Material}
\end{center}
\twocolumngrid
\setcounter{equation}{0}
\setcounter{section}{0}
\setcounter{figure}{0}
\setcounter{table}{0}
\setcounter{page}{1}

\renewcommand{\theequation}{S\arabic{equation}}
\renewcommand{\thefigure}{S\arabic{figure}}
\renewcommand{\thesection}{S\arabic{section}}
\renewcommand{\bibnumfmt}[1]{[S#1]}

\section{Device fabrication and wiring}
\label{section:supmat_fabrication}
\begin{figure}
    \centering
    \includegraphics[width=0.8\columnwidth]{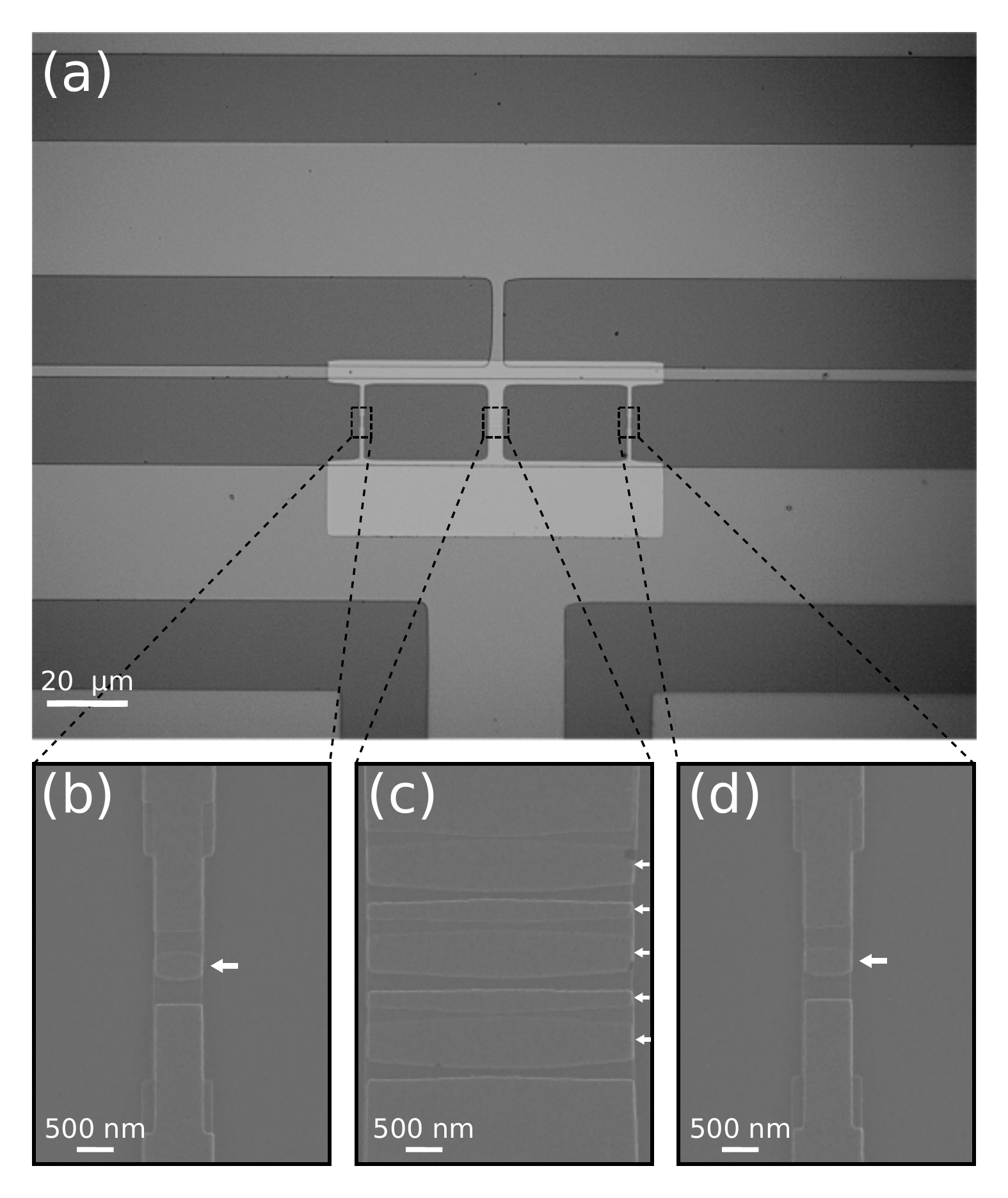}
    \caption{(a) Optical micrograph of the asymmetrically threaded SQUID (ATS) made of aluminium (light grey) deposited on the niobium circuit (grey) over a silicon substrate (dark grey). (b-d) Scanning electron microscope images of the small single junctions (b,d) forming the SQUID loop and the five array junctions (c) forming the inductive shunt. The small junctions are 275 nm $\times$ 700 nm. The array junctions, which would ideally all be equal in area, are in fact composed of three 600nm $\times$ 3.9~µm and two 270 nm $\times$  3.9~µm junctions. This results in a critical current density of about 450 nA$/$µm$^{2}$. For clarity, small arrows point to the location of each junction.}
    \label{fig:FigJJ}
\end{figure}

\paragraph{Wafer preparation} 
We sputter 120~nm of Nb on a 2-inch intrinsic silicon wafer, with a 280~µm thickness and a resistivity larger than 10~k$\Omega$cm. We fabricate twelve 10$\times$11 mm chips on the same wafer. We dice the individual chips at the end of the fabrication process and select the sample that is best suited for the experiment.
\paragraph{Circuit patterning} 
We pattern the large features of the circuit ($>5$ µm) using laser lithography. We spin positive resist (S1805), expose the pattern, then develop in MF319 for 1 min and rinse in deionized (DI) water for 1 min. The wafer is then etched in a reactive ion etching machine with a SF6 plasma and a 10 s overetch. A 30 min lift-off step follows in a 50 °C acetone bath with sonication. Finally, the sample is rinsed in isopropyl alcohol (IPA) for 1 min, blow dried and cleaned in an O$_2$ plasma for  20s, thus stripping residual organic contaminants.
\paragraph{Junction patterning}
Our Josephson junctions are fabricated from Dolan bridges patterned with electron beam (e-beam) lithography. We spin two layers of resist: first, methacrylic acid/
methyl methacrylate (MAA EL13) baked for 3 min at 185 °C and second, poly(methyl methacrylate) (PMMA A3) baked for 30 min at 185 °C. Once the e-beam patterning completed, we develop in a IPA:H$_2$O (3:1) bath at 6 °C for 2 min, rinse for 10 s in IPA and blow dry. Finally, residual organic contaminants below the bridges are stripped by an O$_2$ plasma for 10 s.
\paragraph{Junction deposition}
The wafer is then introduced in an e-beam evaporator. We start with a 2 min argon milling step at an angle of $\pm$ 30° to prepare for a good electrical contact with the Nb layer. We deposit two layers of aluminium (35 nm then 70 nm thick) at an angle of $\pm$ 30°, separated by a static oxidation in a pure O$_2$ atmosphere at 10 mbar for 10 min. Before venting to air, the chamber is filled with 300 mbar of O$_2$ for 5 min. We lift-off in a 50 °C acetone bath for 1 h, transfer the wafer to a new acetone bath for 5 min and sonicate for 10 s, then rinse in IPA and blow dry. Images of the fabricated junctions are displayed in Fig.~\ref{fig:FigJJ}.

\begin{figure*}
    \centering
    \includegraphics[width=1.8\columnwidth]{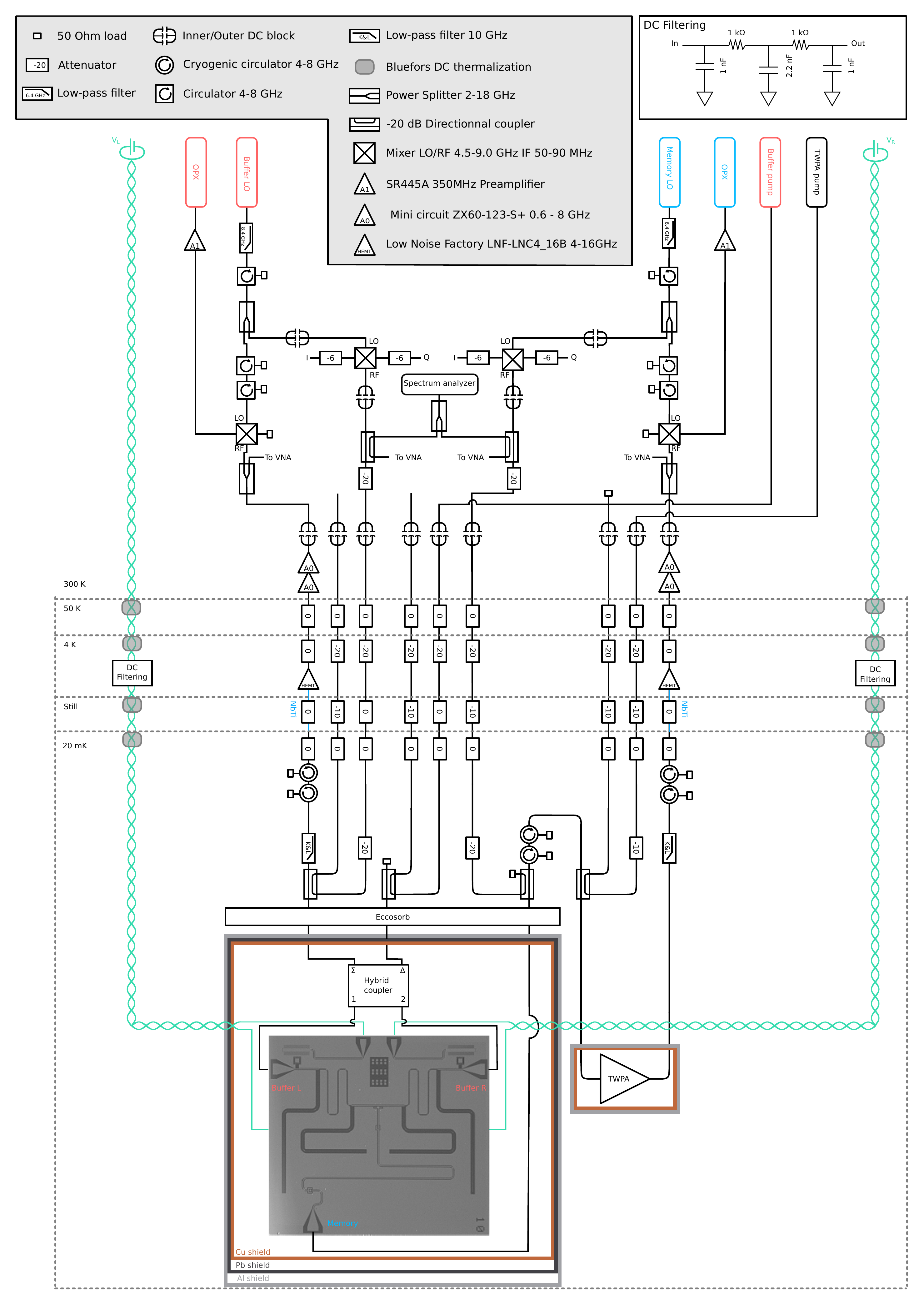}
    \caption{Wiring of the experiment. Measurement apparatus for the memory (blue labels), buffer (red labels), and TWPA pump (black label) connect to the experiment through RF lines (black lines). DC voltage sources are used to drive flux lines (green lines). Dashed lines indicate the different temperature stages of the dilution refrigerator. Additional information is provided in the legend (grey background), annotations and in the text.}
    \label{fig:wiring}
\end{figure*}
\begin{figure}
    \includegraphics[width=0.8\columnwidth]{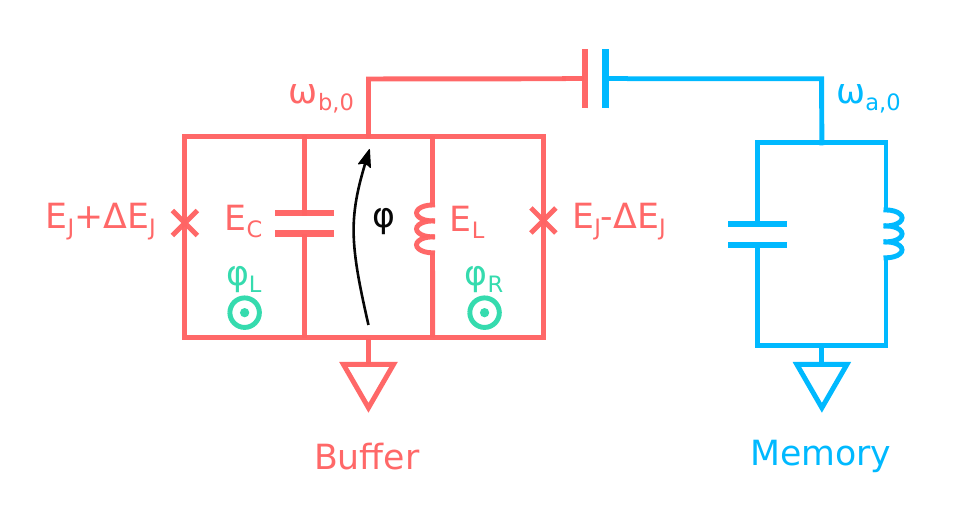}
    \caption{Lumped element model of the circuit. The buffer (red) with bare frequency $\omega_{b,0}/2\pi$ is constituted with an ATS (inductive energy $E_L$, mean Josephson energy $E_J$, asymmetry $\Delta E_J$), and a capacitor with charging energy $E_C$. The ATS loops are threaded with fluxes $\varphi_L$, $\varphi_R$ (green) and is connected to the memory (blue), with bare frequency $\omega_{a,0}/2\pi$. The phase $\varphi$ is indicated with an arrow (black).}
    \label{fig:lumpedelement}
\end{figure}

\section{Circuit Analysis}
\label{section:supmat_analysis}
\subsection{Asymmetrically Threaded SQUID}

The Asymmetrically Threaded SQUID (ATS) is the non-linear inductive dipole that mediates the exchange of pairs of photons between the memory and its environment. This dipole consists of a SQUID split in its center by an inductance at a specific DC flux bias and is represented by the circuit of Fig.~\ref{fig:lumpedelement}.

The inductive energy of the ATS writes \cite{Lescanne2020}
\begin{equation}
\begin{split}
    U_{\varphi_\Sigma, \varphi_\Delta}(\varphi) =\frac{1}{2}E_L\varphi^2&-2E_J\cos(\varphi_\Sigma)\cos(\varphi+\varphi_\Delta)\\&+2\Delta E_J\sin(\varphi_\Sigma)\sin(\varphi+\varphi_\Delta) \,.
\end{split}
\end{equation}
where $E_L$ is the inductive energy of the shunt inductance, $E_J\pm \Delta E_J$ are the Josephson energies of the left and right Josephson junctions respectively, $\varphi$ is the superconducting phase difference across the ATS and $\varphi_{\Sigma,\Delta} = (\varphi_L\pm\varphi_R)/2$ are related to the common and differential flux threading the ATS with $\varphi_{L,R}$ threading the left and right loop of the ATS respectively.
\paragraph{Symmetries}
The ATS potential has the following translational symmetries
\begin{equation}
    U_{\varphi_\Sigma, \varphi_\Delta}(\varphi) = U_{\varphi_\Sigma+\pi, \varphi_\Delta+\pi}(\varphi) = U_{\varphi_\Sigma+\pi, \varphi_\Delta-\pi}(\varphi) \,.
\end{equation}
and an inversion symmetry center at $(\varphi_\Sigma, \varphi_\Delta)=(\pi/2, \pi/2)$ such that
\begin{equation}
    U_{\pi/2+\varphi_\Sigma, \pi/2+\varphi_\Delta}(\varphi) = U_{\pi/2-\varphi_\Sigma, \pi/2-\varphi_\Delta}(-\varphi) 
\end{equation}
Combining these three symmetries gives rise to a second inversion symmetry center located at $(\varphi_\Sigma, \varphi_\Delta)=(\pi/2, -\pi/2)$. Hence, all the information about the system is contained in the region $\varphi_\Sigma\in [0, \pi]$, $\varphi_\Delta\in [-\pi/2, \pi/2]$. 
Note that provided $\Delta E_J=0$, we have additional symmetry axes $\varphi_\Sigma=0$ and $\varphi_\Delta=0$ such that 
\begin{equation}
    U_{\varphi_\Sigma, \varphi_\Delta}(\varphi) = U_{-\varphi_\Sigma, \varphi_\Delta}(\varphi) 
    =U_{\varphi_\Sigma, -\varphi_\Delta}(-\varphi) 
\end{equation}

\paragraph{Saddle points}
Let us study the potential around the inversion symmetry points $(\varphi_\Sigma, \varphi_\Delta)=(\pi/2+\epsilon, \pm\pi/2+\delta)$
\begin{equation}
\begin{split}
    U(\varphi, \epsilon, \delta)=\frac{1}{2}E_L\varphi^2 &\mp 2E_J\sin(\epsilon)\sin(\varphi+\delta)\\&\pm2\Delta E_J\cos(\epsilon)\cos(\varphi+\delta)
\end{split}
\end{equation}

For small $\epsilon$ and $\delta$
\begin{equation}
\begin{split}
    U(\varphi, \epsilon, \delta)&=\frac{1}{2}E_L\varphi^2\\ &
    \mp (- 2\Delta E_J + 2E_J\epsilon\delta + \Delta E_J (\epsilon^2+\delta^2))\cos(\varphi)\\&
    \mp (2E_J\epsilon+ 2\Delta E_J\delta)\sin(\varphi) \,.
\end{split}
\end{equation}

For $\epsilon=\delta=0$, the potential reaches its minimum at $\varphi_\text{min}=0$. At $\epsilon, \delta\ne 0$, we search for a first order perturbation of $\varphi_\text{min}$. Solving for $\frac{\partial}{\partial_\varphi}U(\varphi_\text{min}, \epsilon, \delta)=0$, we get
\begin{equation}
\begin{split}
    \varphi_\text{min}=\pm\frac{2E_J\epsilon+2\Delta E_J\delta}{E_L\mp2\Delta E_J}\,.
\end{split}
\end{equation}
Around the minimum $\varphi_\text{min}$, the second derivative of the potential with respect to $\varphi$, i.e. the inductive energy of the ATS writes
\begin{equation}
\begin{split}
    \frac{\partial^2}{{\partial\varphi}^2}U(\varphi_\text{min}, \epsilon, \delta)&=E_L \mp 2\Delta E_J + E_L\varphi_\text{min}^2\pm 2E_J\epsilon\delta\\
    &\mp \Delta E_J (\varphi_\text{min}^2  -  \epsilon^2-\delta^2)
\end{split}
\end{equation}
The ATS inductive energy has no linear terms in $\epsilon$ or $\delta$ so the points $(\varphi_\Sigma, \varphi_\Delta)=(\pi/2, \pm\pi/2)$ are critical points of the inductive map of the ATS as a function of $\epsilon$ and $\delta$. Its quadratic dependence around the critical point has the following matrix representation
\begin{equation}
\begin{split}
    M(E_L, E_J, \Delta E_J) &= \frac{4(E_L\mp\Delta E_J)}{(E_L\mp2\Delta E_J)^2}\begin{bmatrix} E_J^2 & E_J \Delta E_J \\E_J \Delta E_J & \Delta E_J^2 \end{bmatrix} \\
    &\pm \begin{bmatrix} \Delta E_J & E_J  \\ E_J &  \Delta E_J \end{bmatrix}
\end{split}
\end{equation}
the determinant of which writes
\begin{equation}
    \det(M) =E_L^2\frac{\Delta E_J^2-E_J^2}{(E_L\mp2\Delta E_J)^2}\,.
\end{equation}
The determinant is negative (provided $\Delta E_J < E_J$) hence the critical point is a saddle point. This property is used to tune the DC working point experimentally (see Fig.~\ref{fig:saddlepoint}). When $\Delta E_J \neq 0$, the two points $(\varphi_\Sigma, \varphi_\Delta)=(\pi/2, \pm\pi/2)$ are non equivalent saddle points of the ATS with inductive energy $E_L \mp 2\Delta E_J$.

\subsection{Circuit Hamiltonian}
The dynamics of the circuit displayed in Fig.~\ref{fig:Fig1} is well captured by a reduced lumped element model (see Fig.~\ref{fig:lumpedelement}) with the following Hamiltonian \cite{Lescanne2020}:
\begin{equation}
\label{eq:Htot}
\begin{split}
    H &= \hbar\omega_{a,0} a^\dag a+ \hbar\omega_{b, 0} b^\dag b \\
    &-2E_J \cos(\varphi_\Sigma)\cos(\varphi + \varphi_\Delta)\\
    &+2\Delta E_J \sin(\varphi_\Sigma)\sin(\varphi+\varphi_\Delta)
\end{split}
\end{equation}
where $a,b$ are the memory and buffer annihilation operators. The buffer's angular frequency verifies $\omega_{b,0}=\sqrt{8E_LE_C}/\hbar$, where $E_L, E_C$ are the energies associated to the buffer's inductive and capacitive shunt respectively. The angular frequency of the memory is denoted $\omega_{a,0}$. We denote $2E_J$ the sum of the Josephson energies of the single junctions composing the SQUID loop, and $2\Delta E_J$ their difference. During fabrication we aim for the smallest possible junction asymmetry, however in practice we are left with $\Delta E_J / E_J\approx 0.5\%$ (see \cite{supmat},  Sec.~\ref{sec:DC_currents}) which leads to spurious Kerr and cross-Kerr effects. We neglect $\Delta E_J$ in the rest of the analysis. The ATS is threaded with a common and differential flux $\varphi_{\Sigma,\Delta}=\frac{1}{2}(\varphi_L\pm\varphi_R)$, where $\varphi_{L,R}$ is the flux threading the left and right loop of the ATS. In the limit where the hybridization factor $\upsilon$ between the buffer and memory is much smaller than 1, the phase across the ATS denoted $\varphi$, verifies $\varphi = \varphi_b \left(b+b^\dag+\upsilon(a+a^\dag)\right)$, where the zero point phase fluctuations for the buffer reads $\varphi_{b}=\left(2E_C/E_L\right)^{1/4}$.

\subsection{First order Hamiltonian at the operating point}
By flux pumping the ATS around a well chosen DC working point \cite{Lescanne2020}
\begin{equation}
    \begin{split}
        \varphi_\Sigma &= \frac{\pi}{2} + \epsilon_p \cos(\omega_p t) \\
        \varphi_\Delta &= \frac{\pi}{2}
    \end{split}
\end{equation}
we engineer a two-to-one photon exchange Hamiltonian between the memory and the buffer, provided the pump frequency $\omega_p$ is close to the matching condition $\omega_p= 2\omega_a - \omega_b$. This two-to-one photon exchange Hamiltonian converts the strong single photon losses of the buffer into an effective two-photon loss channel for the memory. 

Likewise, a microwave drive at frequency $\omega_d$ close to the buffer frequency, is converted into an effective two-photon drive of the memory (or squeezing) at frequency $(\omega_d+\omega_p)/2$. By definition, this frequency is close to the memory frequency.

For the memory, the combination of the two-photon loss and two-photon drive, stabilizes two coherent states with frequency $(\omega_p+\omega_d)/2$ of equal amplitude and opposite phase. The heterodyne demodulation frequency $\omega_\text{dm}$ for the memory is constrained accordingly 
\begin{equation}
        \omega_\text{dm} = \frac{\omega_p+\omega_d}{2}     \,.
\end{equation}

By going in the frame rotating at frequency $\omega_\text{dm}$ for the memory and $\omega_d$ for the buffer and performing first order rotating wave approximation (RWA), the Hamiltonian \eqref{eq:Htot} writes \cite{Lescanne2020}
\begin{equation}
        H/\hbar = -\Delta_a a^\dag a -\Delta_b b^\dag b + g_2^*a^2 b^\dag + g_2{a^2}^\dag b
\end{equation}
where $\Delta_a = \omega_\text{dm} - \omega_a$, $\Delta_b = \omega_d -\omega_b$ and $\omega_a$ and $\omega_b$ are respectively the memory and buffer frequency accounting for the AC-stark shift due to the pump \cite{Lescanne2020}. 

Incorporating the buffer drive and the dissipation of the two modes, the dynamics of the system is governed by 
\begin{equation}
    \begin{split}
        H/\hbar &=-\Delta_a a^\dag a - \Delta_b b^\dag b \\ &\qquad+g_2^*a^2{b}^\dag + g_2a^{\dag 2}{b}-\epsilon_d b^\dag-\epsilon_d^* b\\
        L_a &= \sqrt{\kappa_a}a \\
        L_b &= \sqrt{\kappa_b}b
    \end{split}
    \label{eq:hrot}
\end{equation}

where $\epsilon_d$ is the buffer drive strength and  $\kappa_a$ and $\kappa_b$ are the single photon loss rate of the memory and the buffer respectively. 

We gain further insight on the dynamics of the system by performing the adiabatic elimination of the buffer. This is justified provided $g_2 \ll \kappa_b$. Following the method of \cite{Azouit2016}, we find that the reduced dynamics of the memory is given by 
%
\begin{equation}
\label{eq:adiab_elim}
    \begin{split}
        H_a/\hbar &= -\Delta_a a^\dag a - \epsilon_d \gamma a^{\dag2} - \epsilon_d \gamma^* a^2 + \Delta_b|\gamma|^2 a^{\dag2}a^2 \\
        &L_{a^2}=\sqrt{\kappa_b|\gamma|^2} a^2 \\
        &L_a = \sqrt{\kappa_a}a
    \end{split}
\end{equation}
\begin{equation}
    \text{with} \quad \gamma = \frac{g_2}{\Delta_b+i\kappa_b/2}     \,.
\end{equation}
At $\Delta_a=\Delta_b=0$, we recover eq.~\eqref{eq:L2L1}
\begin{equation}
\label{eq:nodet_dynamics}
\begin{split}
H_2/\hbar&=i\epsilon_2a^{\dag2}-i\epsilon_2^*a^2 \\
&L_2 =\sqrt{\kappa_2}a^2 \\ &L_1 = \sqrt{\kappa_a}a
\end{split}
\end{equation}
with $\epsilon_2 = 2\epsilon_d g_2/\kappa_b$ and $\kappa_2 = 4|g_2|^2/\kappa_b$. 

\subsection{Steady-state photon number}
\label{section:steadystatephotonnumber}

In this paragraph, we derive the stationary  mean photon number in the memory using a semi-classical
approximation

In the interaction picture, the dynamics arising from \eqref{eq:hrot} writes
\begin{equation}
    \begin{split}
        \frac{da}{dt} &= \left( i\Delta_a - \frac{\kappa_a}{2} \right) a -2i g_2 a^\dag b \\
        \frac{db}{dt} &= \left( i\Delta_b - \frac{\kappa_b}{2} \right) b - i g_2^* a^2 + i\epsilon_d   \,.
    \end{split} 
\label{eq:eom}
\end{equation}

We perform a mean-field approximation on mode $a$ and $b$, and compute the steady-state of the simplified dynamics. The operator $a$ and $b$ are replaced by their mean value, the complex numbers $\alpha$ and $\beta$.

This system always admits a solution in which the memory is in vacuum and corresponds to
\begin{equation}
\begin{split}
    \alpha &=0 \\
    \beta &= \frac{-\epsilon_d}{\Delta_b+i\kappa_b/2}\,.
\end{split}
\end{equation}
This solution is stable provided it is the only solution of eq.~\eqref{eq:eom} for a given set parameters.

Assuming $\alpha \neq 0$, we can write
\begin{equation}
\begin{split}
    i\frac{\kappa_a}{2}+\Delta_a &= 2 g_2 \beta e^{-2i\theta_a} \\
    \left( i\frac{\kappa_b}{2}+\Delta_b \right) \beta &= g_2^* \alpha^2 - \epsilon_d
\end{split}
\label{eq:42}
\end{equation}
where $\theta_a=\arg(\alpha)$.

Solving for $\beta$ in the first equation and injecting in the second one, we get
\begin{equation}
\begin{split}
|\alpha|^2 &= \frac{\epsilon_d}{g_2^*} e^{-2i\theta_a} + z \\
\text{with}\quad z &= \frac{(i\kappa_a/2+\Delta_a)(i\kappa_b/2+\Delta_b)}{2 |g_2|^2}\,.
\label{eq:intersection}
\end{split}
\end{equation}

\paragraph{Zero-detuning}
When $\Delta_a =\Delta_b=0$, eq.~\eqref{eq:intersection} simplifies into 
\begin{equation}
\label{eq:intersectionzero}
    |\alpha|^2 = \frac{\epsilon_d}{g_2^*} e^{-2i\theta_a} - \frac{\kappa_a\kappa_b}{8|g_2|^2}\,.
\end{equation}
leading to 
\begin{equation}
\label{eq:nbar}
\begin{split}
    |\alpha|^2 &= \max\left[\left|\frac{\epsilon_d}{g_2^*}\right|- \frac{\kappa_a\kappa_b}{8|g_2|^2}, 0\right] \\ 
    &=
    \max\left[\left|\frac{\epsilon_d}{g_2^*}\right|\left(1 - \frac{\kappa_a}{4|\epsilon_2|}\right), 0\right]     \,.
\end{split}
\end{equation}
We recover the critical point when the two-photon drive overcomes the cavity dissipation.

In the absence of calibrated input or output lines, the power radiated by the memory is defined up to a constant, in particular the quantity $|g_2\alpha|^2$ writes
\begin{equation}
\label{eq:nbarg22}
    |g_2\alpha|^2 = \max\left[
    |\epsilon_d g_2| -  \frac{\kappa_a\kappa_b}{8}, 0\right]
\end{equation}
which, as a function of $\epsilon_d g_2$, has a slope 1 and an x-intercept $\kappa_a\kappa_b/8 \,.$

In terms of the single mode effective quantities, eq.~\eqref{eq:nbar} rewrites for $\epsilon_2\ge \kappa_a/4$ 
$$|\alpha|^2 = \frac{1}{2\kappa_2}(4\epsilon_2-\kappa_a)\qquad\text{(two-photon dissipation)}\;.$$ 
If instead of two-photon dissipation, Kerr effect of amplitude $K$ was limiting the amplitude of the pointer states \cite{Bartolo2016, Puri2017}, a similar semi-classical analysis predicts a mean photon number for $\epsilon_2\ge \kappa_a/4$
$$|\alpha|^2 = \frac{1}{2K}\sqrt{(4\epsilon_2)^2-(\kappa_a)^2}\qquad\text{(dissipative Kerr)}\;,$$
which is qualitatively different from what is observed in this experiment.

\paragraph{General case}
Since $\theta_a$ and $\alpha$ are on separate sides of eq.~\eqref{eq:intersection}, we can  geometrically solve the system in the complex plane. The right-hand side is a circle of radius $|\epsilon_d/g_2^*|$ centered on $z$. The left-hand side is the positive real axis. In this picture, there can be 0, 1, or 2 intersections  between this circle and the real positive axis, giving rise to 1, 3 or 5 solutions for the system, one for vacuum plus two for each intersection since $\pm \alpha$ are both valid solutions. Experimentally observed solutions are the ones that give rise to the largest field in the memory. Hence $|\alpha|^2$, the mean photon number in steady-state, writes:

\begin{equation}
\label{eq:nbardet}
    \!\!|\alpha|^2 =
    \begin{cases}
    \max\!\left[ \text{Re}(z)\!+\! \sqrt{\left|\frac{\epsilon_d}{g_2^*}\right|^2\!\!-\text{Im}(z)^2}, 0 \right]\!\!,& \!\!\!\text{if } \left|\frac{\epsilon_d}{g_2^*}\right|^2\!\!\!>\!\text{Im}(z)^2\\
    0\,,              & \!\! \!\text{otherwise.}
    \end{cases}
\end{equation}

In the $\Delta_a, \Delta_b$ coordinates, the region where $\alpha^2$ is non-zero forms a diamond shape. Here after, we provide the equation for the borders of this feature colloquially referred to as a diamond. 

The $\Delta_a, \Delta_b$-plane is divided in two domains depending on the sign of the quantity $\Delta_a \Delta_b - \kappa_a\kappa_b/4 \,.$

The top-right and bottom-left edges of the diamonds are located in the positive domain and given by
\begin{equation}
    \kappa_a \Delta_b + \kappa_b \Delta_a = \pm 4|\epsilon_d g_2|\,.
\label{eq:diamond_edge}
\end{equation}
Note that the slope of this edge is given by $-\kappa_b/\kappa_a$.
The bottom-right and top-left edges are located in the negative domain and given by
\begin{equation}
    (\frac{\kappa_a^2}{4}+\Delta_a^2)(\frac{\kappa_b^2}{4}+\Delta_b^2)=|2\epsilon_d g_2|^2\,.
\end{equation}
Hence the border of the diamond only depends on the product $\epsilon_d g_2$ and does not carry information on $g_2$ nor $\epsilon_d$ independently. 
Moreover, we cannot determine $g_2$ and $\epsilon_d$ even with the full diamond information (not only the edges). Indeed, from the measurement of the rates $\kappa_a$, $\kappa_b$ and the knowledge of the applied detunings $\Delta_a$, $\Delta_b$, one has independently access to the quantity 
\begin{equation}
    z' =\frac{1}{2} (i\frac{\kappa_a}{2}+\Delta_a)(i\frac{\kappa_b}{2}+\Delta_b)\,.
\end{equation}
In the absence of photon number calibration, we only learn from eq.~\ref{eq:nbardet} that the power radiated by the memory is proportional to
\begin{equation}
\label{eq:powerdet}
    \braket{I^2} \propto
    \text{Re}(z') + \sqrt{|\epsilon_d g_2|^2-\text{Im}(z')^2}\,.
\end{equation}
Thus, we only have access to the product $\epsilon_d g_2$ but not $g_2$ or $\epsilon_d$ independently.


\section{Tuning the experiment}
\label{section:supmat_tuning_cat_qubit}
Our experiment requires DC and RF powering. For optimal operation, two DC currents and two RF powers and frequencies need to be fine tuned. In this section we describe the sequence of calibration experiments we perform to set the working point.

\subsection{DC currents}
\label{sec:DC_currents}
\begin{figure}
\includegraphics[width=1\columnwidth]{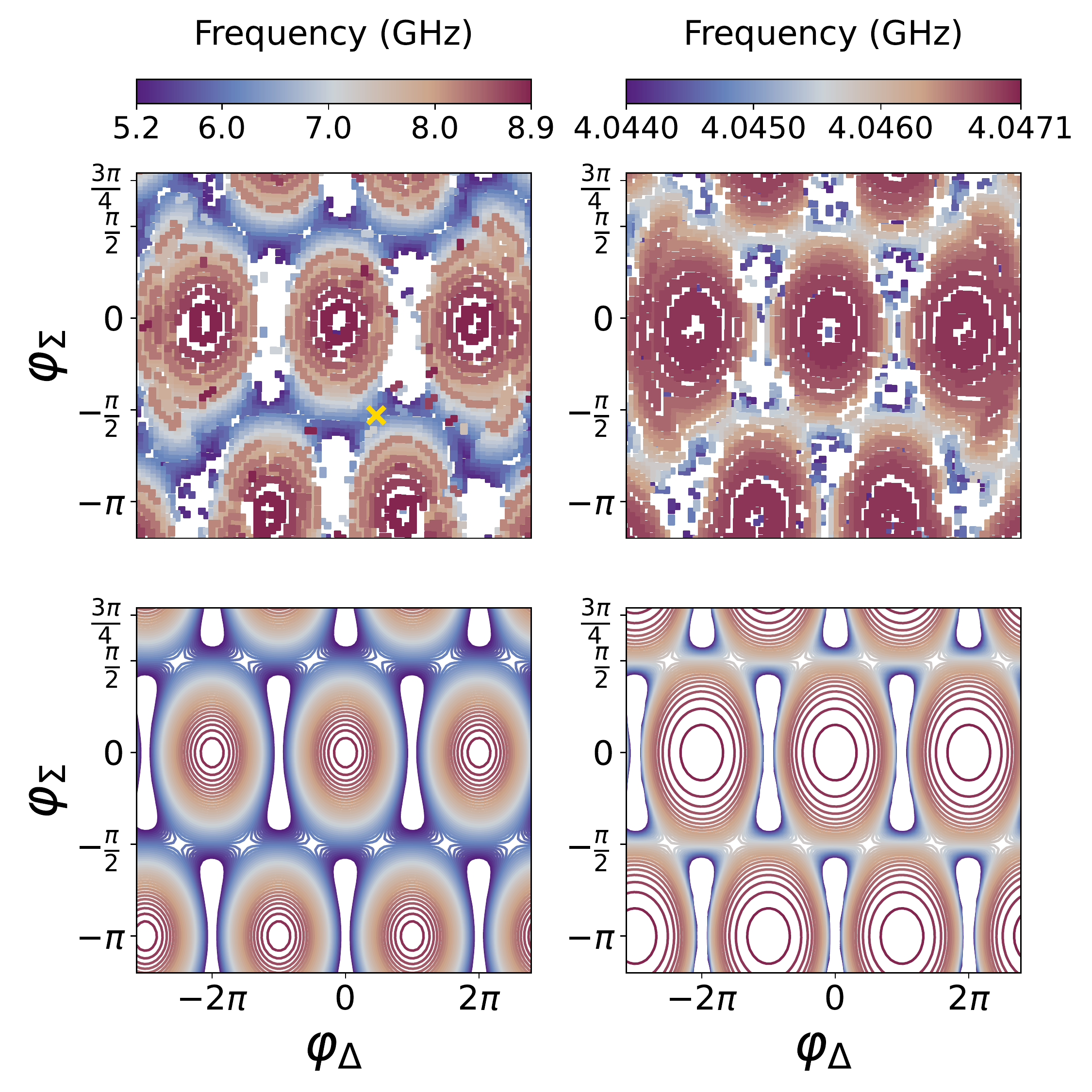}
\caption{Measured (top) and simulated (bottom) frequencies (color) of the buffer (left) and memory (right) as a function of the differential flux (x-axis) and common flux (y-axis) in the ATS loop. The orange cross marks the flux point at which we operate the experiment.}
\label{fig:fluxmap}
\end{figure}

The first calibration experiment we perform is to extract the buffer and memory frequencies as a function of the common and differential flux in the ATS loop (see Fig.~\ref{fig:fluxmap}). From these maps we identify the circuit parameters and locate the ATS saddle points.

In the following we describe the measurement protocol to acquire the buffer frequency flux map. We set a tone at frequency $f$ on the buffer port and record its reflected amplitude and phase as a function of the DC voltages $V_{\Sigma,\Delta}=(V_L\pm V_R)/2$ (see Fig.~\ref{fig:wiring}). The physical controls $V_{\Sigma, \Delta}$ are transformed to the common and differential flux basis $\varphi_{\Sigma,\Delta}$ to match the symmetries of the circuit Hamiltonian \eqref{eq:Htot}. A variation in the reflected signal is detected at flux points $\varphi_{\Sigma,\Delta}(f)$ where the buffer frequency enters the vicinity of $f$.
This sequence is repeated by scanning $f$ in between 5.2 GHz and 9 GHz in steps of 100 MHz. In Fig.~\ref{fig:fluxmap}, the frequency $f$ is encoded in the color of pixels located at $\varphi_{\Sigma,\Delta}(f)$. We repeat the same protocol on the memory port to extract the memory frequency flux map.

The theory plots in Fig~\ref{fig:fluxmap} are obtained for the numerical diagonalization of the Hamiltonian in Eq.~\eqref{eq:Htot}. From the ATS symmetries, we know that there exist two nonequivalent families of saddle points, those generated from ${(\varphi_\Delta, \varphi_\Sigma)=(-\pi/2, \pi/2)}$, and those from ${(\varphi_\Delta, \varphi_\Sigma)=(\pi/2, \pi/2)}$. The junction asymmetry $\Delta E_J$ lifts the degeneracy of the buffer frequency between these points.

\begin{figure}
\centering
\includegraphics[width=1\columnwidth]{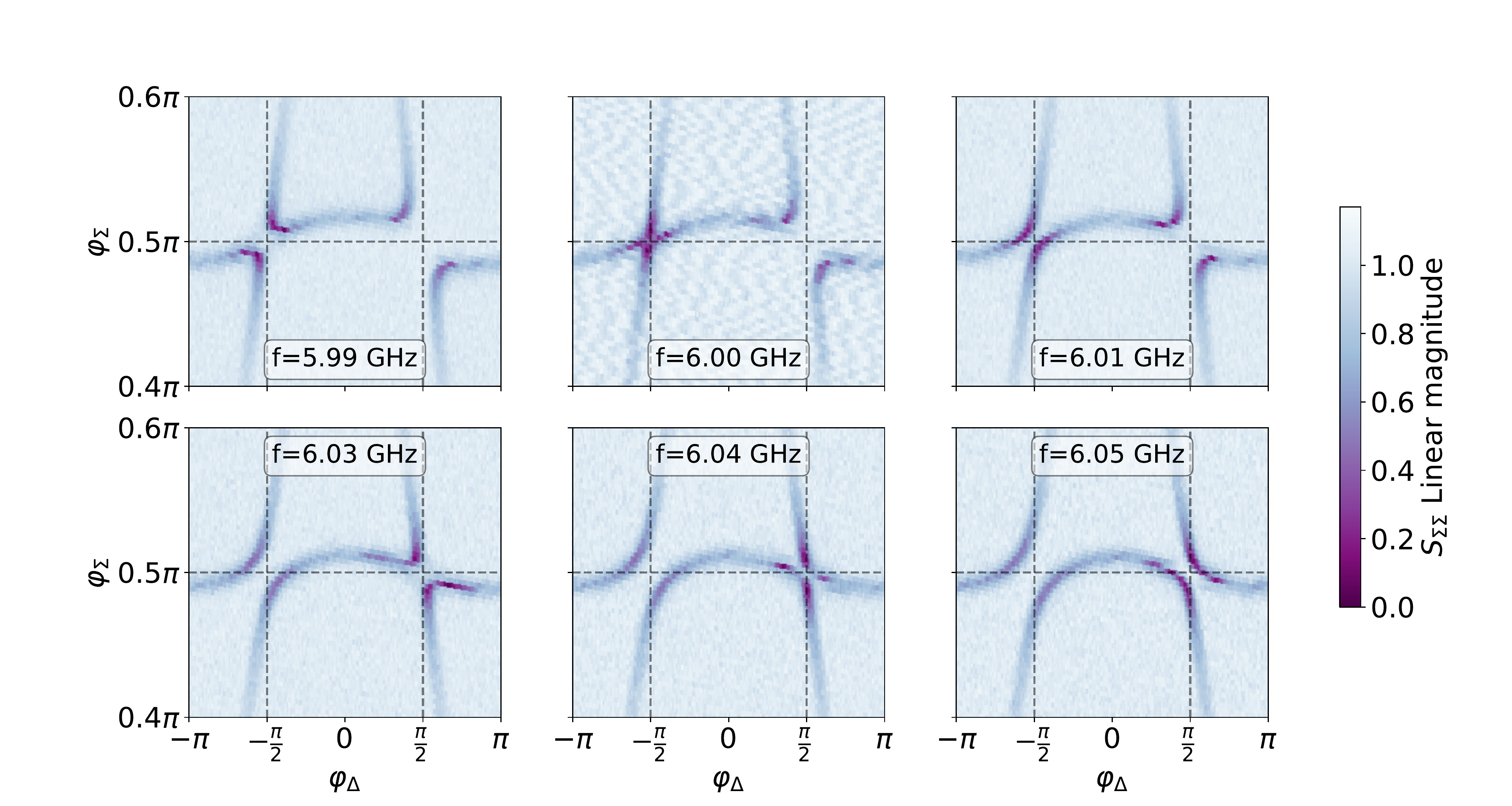}
\caption{Each panel displays the measured relative amplitude (color) of the reflected signal on the buffer port at the frequency $f$ specified in each label box, as a function of the differential (x-axis) and common (y-axis) phase biases. The dashed grey lines are guides for the eye corresponding to $\varphi_{\Delta}=\pm \pi/2$ and $\varphi_\Sigma=\pi/2$.  \label{fig:saddlepoint}}
\end{figure}

We refine the flux and frequency sweeps around these saddle points in order to precisely pin down their location. In Fig.~\ref{fig:saddlepoint}, we directly display the reflected amplitude on the buffer port at different frequencies $f$. A saddle point is easily identified as the closing of the buffer frequency contour line. Note that the saddle point at $(-\pi/2, \pi/2)$ appears at $f_{b1}=6.00 \text{ GHz}$ as shown in the top middle panel. The second one appears at $f_{b2}=6.04 \text{ GHz}$ as shown in the bottom middle panel.

The parameters entering Eq.~\eqref{eq:Htot} are listed in table \ref{tab:params}. The charging energy $E_C$ is extracted from 3D finite elements electromagnetic simulations. The inductive energy $E_L$ and the junction asymmetry $\Delta E_J$ are computed from the buffer frequencies at the saddle points that verify, in the weakly hybridized limit:
$$
f_{b1,b2}=\frac{1}{h}\sqrt{8 E_C (E_L\pm 2 \Delta E_J)}\;.
$$
The Josephson energy $E_J$ is extracted from the maximum buffer frequency $f_{b_\text{max}}$ measured in the buffer flux map of Fig.~\ref{fig:fluxmap}. This maximum value $f_{b_\text{max}}=8.9~\text{GHz}$ is reached for ${(\varphi_\Delta, \varphi_\Sigma)=(0, 0)}$ and verifies, in the weakly hybridized limit: ${f_{b_\text{max}}=\frac{1}{h}\sqrt{8 E_C (E_L+2 E_J)}}$.

The memory frequency $f_a$ is extracted from the memory frequency flux map in Fig.~\ref{fig:fluxmap} at the saddle points. Due to the weak hybridization with the buffer, the memory frequency difference at the two saddle points is negligible. Finally we numerically find the hybridization factor $\upsilon$ that produces a memory frequency flux map in agreement with the data in Fig.~\ref{fig:fluxmap}.

\begin{center}
\begin{table}
\begin{tabular}{ |c|c| } 
 \hline
$E_C/h$ & 72.6 MHz\\ 
\hline
$E_L/h$ & 62.40 GHz \\ 
 \hline
$E_J/h$ & 37.00 GHz \\ 
 \hline
 $\Delta E_J/h$ & 0.207 GHz \\ 
 \hline
 $\omega_{b,0}/2\pi$ & 6.020 GHz\\ 
 \hline
  $\varphi_{b}$ & 0.220 \\
  \hline
   \hline
$\omega_{a,0}/2\pi$ & 4.0457 GHz\\ 
\hline
  $\upsilon$ & 3.6\% \\
  \hline
\end{tabular}

\caption{Buffer and memory parameters entering the Hamiltonian \eqref{eq:Htot}. $E_c$ and $E_l$ are respectively the buffer's charging and inductive energy. $E_J$ and $\Delta E_J$ are the ATS mean Josephson energy and asymmetry. $\omega_{b, 0}/2\pi$ and $\omega_{a, 0}/2\pi$ are buffer and memory bear frequencies. $\varphi_b$ is the buffer zero point fluctuations. $\upsilon$ is the hybridization strength. From these numbers, we can estimate the Kerr non-linearity of the memory $K_a<1$ Hz.}
\label{tab:params}
\end{table}
\end{center}

\begin{center}

\begin{table}[]
    \centering
    \begin{tabular}{|c|c|}
     \hline
          $\omega_b/2\pi$ & 6.1273 GHz\\ 
     \hline 
          $\omega_a/2\pi$ & 4.0458 GHz\\ 
     \hline 
        $\omega_p/2\pi$ & 2.07 GHz\\ 
     \hline
    \end{tabular}
    \begin{tabular}{|c|c|}
    \hline
    $\kappa^i_a/2\pi$ & 18 kHz\\ 
    \hline
    $\kappa^c_a/2\pi$ & 40 kHz\\ 
    \hline
    $\kappa_b/2\pi$ & 16 MHz\\
    \hline
    \end{tabular}
    \begin{tabular}{|c|c|}
    \hline
    $g_2/2\pi$ & 39 kHz\\ 
    \hline
    $\kappa_2/2\pi$ & 370 Hz\\ 
    \hline
    $\overline{n}_\text{sat}$ & 43\\ 
    \hline
      \end{tabular}
    \caption{Parameters at the operating point of the experiment. $\omega_b/2\pi$, $\omega_a/2\pi$, $\omega_p/2\pi$ are the buffer, memory and pump frequencies. $\kappa_a^i$,  $\kappa_a^c$ are the internal and coupling loss rates of the memory, $\kappa_b$ is the loss rate of the buffer, $g_2$ is the two-photon coupling rate, $\kappa_2/2\pi$ is the two-photon dissipation rate and $\bar n_{\rm sat}$ is the average number of photons at the bit-flip time saturation.
    The following parameters are given with confidence intervals: $\kappa_a^i \in [15,22]$ kHz, $\kappa_a^c \in [39,42]$ kHz, $\kappa_b^i \in [13,20]$ MHz, $g_2/2\pi \in [30, 46]$ MHz,  $\kappa_2 \in [270, 410]$, $\overline{n}_\text{sat} \in [43, 54]$.}
    \label{tab:wkpt_params}
\end{table}
\end{center}

\subsection{Phase-locking}
In the laboratory frame, at any point in time, the phase of the pointer states resulting from the junction mixing process is given by 
\begin{equation}
    \theta_a=(\theta_p+\theta_d)/2
\end{equation}
where $\theta_p$ and $\theta_d$ are respectively the pump and drive tone phase. The pump tone is directly generated by the microwave signal generator and pulsed via a microwave switch whereas the drive tone is pulsed via an IQ-mixer (Fig.~\ref{fig:wiring}).The resulting phases of the tones are
\begin{equation}
    \theta_p= \theta_p^{LO}, \quad \theta_d = \theta_d^{LO}+\theta_d^{IF}
\end{equation}
where $\theta_p^{LO}$ and $\theta_d^{LO}$ are the Local Oscillator (LO) phases of the microwave generator and $\theta_d^{IF}$ is the Intermediate Frequency (IF) signal phase delivered by the Arbitrary Waveform Generator (AWG) channel to pulse the drive tone. 
The radiated signal from the memory is demodulated in a frame with phase
\begin{equation}
    \theta_{dm} = \theta_{dm}^{LO} +\theta_{dm}^{IF}
\end{equation}
where $\theta_{dm}^{LO}$ is the phase of the demodulation LO and $\theta_{dm}^{IF}$ is the phase of the Analog-Digital Converter (ADC). 
In order to phase-lock the pointer states with the demodulation frame, we should ensure 
\begin{equation}
    \theta_a - \theta_{dm} = {\rm cst}
\end{equation}
The three LOs are generated with a single four channel Anapico signal generators, and the two IFs with a single Quantum Machines OPX. The accuracy of these instruments ensure that all the LOs share the same time reference and all the IFs share the same time reference. However, given the high frequencies at stake, the instrument sharing the same 50 MHz clock is not sufficient for this two time references to be considered identical. The LO time is referred to as $t$ and the IF time as $t'$. 
Hence 
\begin{equation}
\begin{split}
    \theta_a - \theta_{dm} &= (\omega_p^{LO}t+\omega_d^{LO}t+\omega_d^{IF}t')/2 + {\rm cst} \\&\quad- (\omega_{dm}^{LO}t+\omega_{dm}^{IF}t') + {\rm cst} \\
    &= ((\omega_p^{LO}+\omega_d^{LO})/2-\omega_{dm}^{LO})t \\
    & \quad+ (\omega_d^{IF}/2-\omega_{dm}^{IF})t' \,.
\end{split}
\end{equation}
and the phase-locking condition imposes the frequency matching conditions 
\begin{equation}
\label{eq:phaselock}
\begin{split}
    (\omega_p^{LO}+\omega_d^{LO})/2-\omega_{dm}^{LO} = 0 \\
    \omega_d^{IF}/2-\omega_{dm}^{IF} = 0
\end{split}
\end{equation}
\subsection{Pump and drive frequencies}

\begin{figure*}
    \centering
    \includegraphics[width=2\columnwidth]{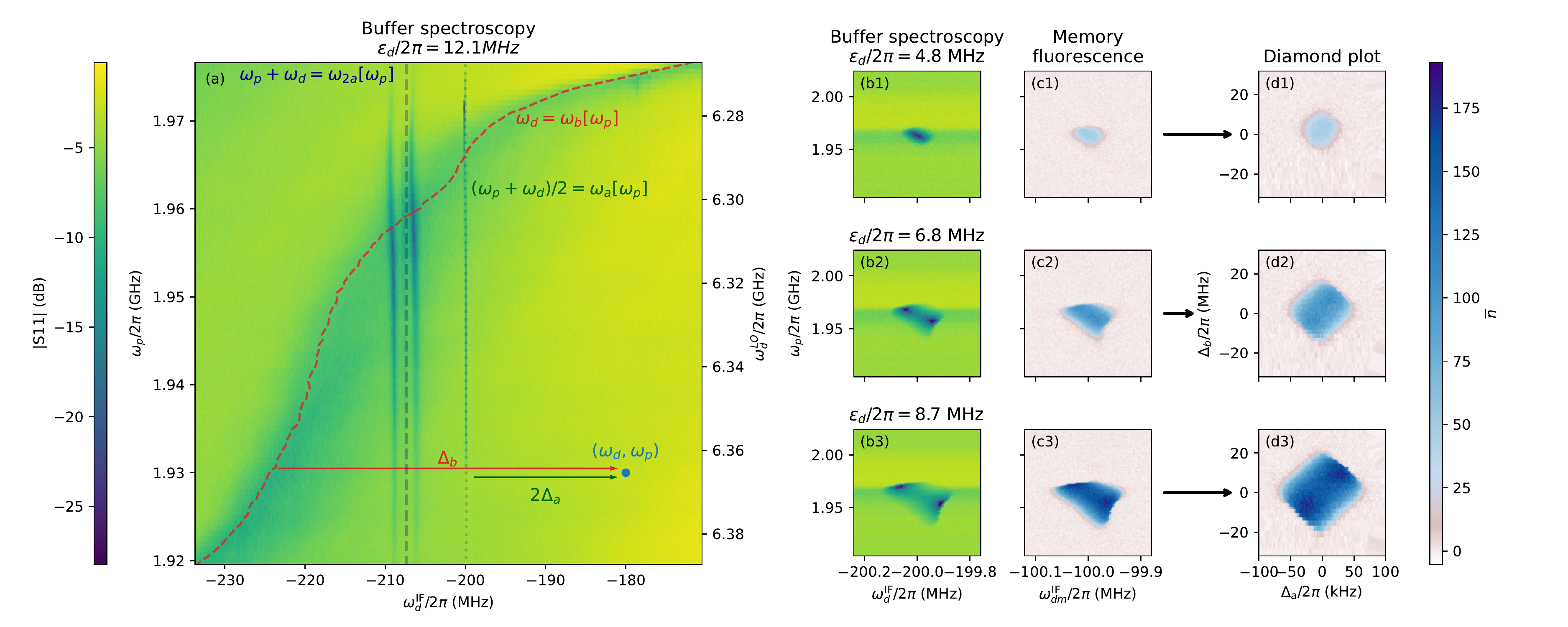}  
    \caption{(\textbf{a}) Relative amplitude (color) of the reflected signal on the buffer port as a function of the pump frequency $\omega_p$ (left y-axis), drive local oscillator (LO) frequency $\omega_d^\text{LO}$ (right y-axis) and drive intermediate frequency $\omega_d^\text{IF}$ (x-axis). The drive frequency is given by $\omega_d = \omega^\text{IF}_d+\omega^\text{LO}_d$. For each pump frequency, the drive LO frequency is set to $\omega^\text{LO}_d=(2\omega_a-\omega_p)+200~\text{MHz}$, such that vertical lines correspond to constant detuning from the frequency matching condition ($\Delta_a = cte$). The buffer drive resonance condition $\Delta_b = 0$ is determined by fitting each horizontal cuts of the map (dashed red line). In the vicinity of $\Delta_a = 0$ and $\Delta_b=0$, a sharp feature indicates that the two-to-one photon exchange transition is resonant (dotted green line). Spurious transitions appear near the frequency matching condition $\omega_p + \omega_d= \omega_{2a}$ (blue dashed line), where $\omega_{2a}$ is the frequency of the second harmonic of the memory $\lambda/2$-resonator measured independently. (\textbf{b}) Zoom on the two-to-one photons exchange  transition for increasing drive amplitude $\epsilon_d$. (\textbf{c}) radiated energy from the memory in units of circulating photon number (color) as a function the pump frequency $\omega_p$ (left y-axis) and two-photon drive intermediate frequency $\omega_{dm}^\text{IF}$ (x-axis), for increasing drive amplitude. On these panels, the two-photon drive LO frequency is set to $\omega^\text{LO}_{dm}=\omega_a+100~\text{MHz}$. When the two-to-one photon exchange transition is resonant, the engineered two-photon drive populates the memory. The average occupation of the memory is determined thanks to an undercoupled port via heterodyne detection. (\textbf{d}) radiated energy from the memory in units of circulating photon number (color) as a function of the pump and drive detuning from the frequency matching condition $\Delta_a$ (x-axis), and the drive detuning from the buffer $\Delta_b$ (y-axis). In these coordinates, the feature takes the shape of a regular diamond.}

    \label{fig:largediamond}
\end{figure*}

Once the DC biases are tuned at one saddle point and the memory and buffer resonance frequencies are determined by direct spectroscopy in reflection on their respective ports, we proceed to tune the pump and drive tones. The pump frequency is determined via two-tone spectroscopy: a weak drive tone is used to perform buffer spectroscopy while sweeping the pump frequency around the frequency matching condition $\omega_p = 2\omega_a-\omega_b$. As this operation is being performed, we also perform the heterodyne detection of the field radiated by the memory (Fig.~\ref{fig:largediamond}).

When the two-to-one photon exchange is resonant, a sharp feature is observed within buffer resonance, referred to as a diamond, and the memory starts to radiate power. The discrepancy between the ideal and measured diamond shape is used as a witness for the appearance of higher order processes. The pump amplitude is set so as to maximize $g_2$ while mitigating detrimental higher order effects.

In order to minimize the amount of data collected and accurately zoom on the diamond feature Fig.~\ref{fig:largediamond}a is acquired in the following way :
\begin{itemize}
    \item at a fixed pump frequency (x-axis of Fig.~\ref{fig:largediamond}) the buffer spectroscopy is done by varying the buffer IF frequency with a fixed LO frequency. Due to the frequency constraint of eq.~\eqref{eq:phaselock} the heterodyne detection of the radiated memory field is done by varying the memory IF frequency with a fixed LO frequency. 
    \item for the demodulation frequency to remain close to the memory frequency while varying the pump frequency, (y-axis of Fig.~\ref{fig:largediamond}) the pump and drive LO frequencies are varied in opposite directions. In this way, we have  $\omega_p+\omega_b^{LO}=2\omega_{dm}^{LO}=\text{cst.}$
\end{itemize}
In these coordinates, $\Delta_b$ and $2\Delta_a$ are varied along the x-axis and $\Delta_b$ is varied along the y-axis. If there was no Stark shift, the buffer resonance would be a diagonal line of slope 1 and the memory line (when the two-photon drive is tuned with the memory mode) a vertical line. In practice, these two lines are distorted and we numerically fit the buffer and memory frequencies as a function of the pump frequency to perform the change of basis leading to the diamond of Fig.~\ref{fig:largediamond}d in the $\Delta_a, \Delta_b$ coordinate system.

To perform this change of coordinates, we evaluate the following functions from measurements by linear interpolation:
\begin{equation}
\begin{split}
    \Delta_b = f(\omega^{IF}_d, \omega_p) &= \omega_d^{IF} + \omega_d^{LO} - \omega_b[\omega_p] \\
    \Delta_a = g(\omega^{IF}_d, \omega_p) &= \frac{\omega_b^{IF}}{2} +  \frac{\omega_p+\omega_d^{LO}}{2} - \omega_a[\omega_p] \\
    &= \omega_{dm}^{IF} + \omega_{dm}^{LO} - \omega_a[\omega_p] \,.
\end{split}
\end{equation}
For each data point, we can now compute the actual value of $\Delta_a$ and $\Delta_b$. This enables us to display radiated energy by the memory in the basis of Hamiltonian eq.~\eqref{eq:hrot}: previously distorted, the diamonds recover their shape. 

\begin{figure}
    \centering
    \includegraphics[width=0.72\columnwidth]{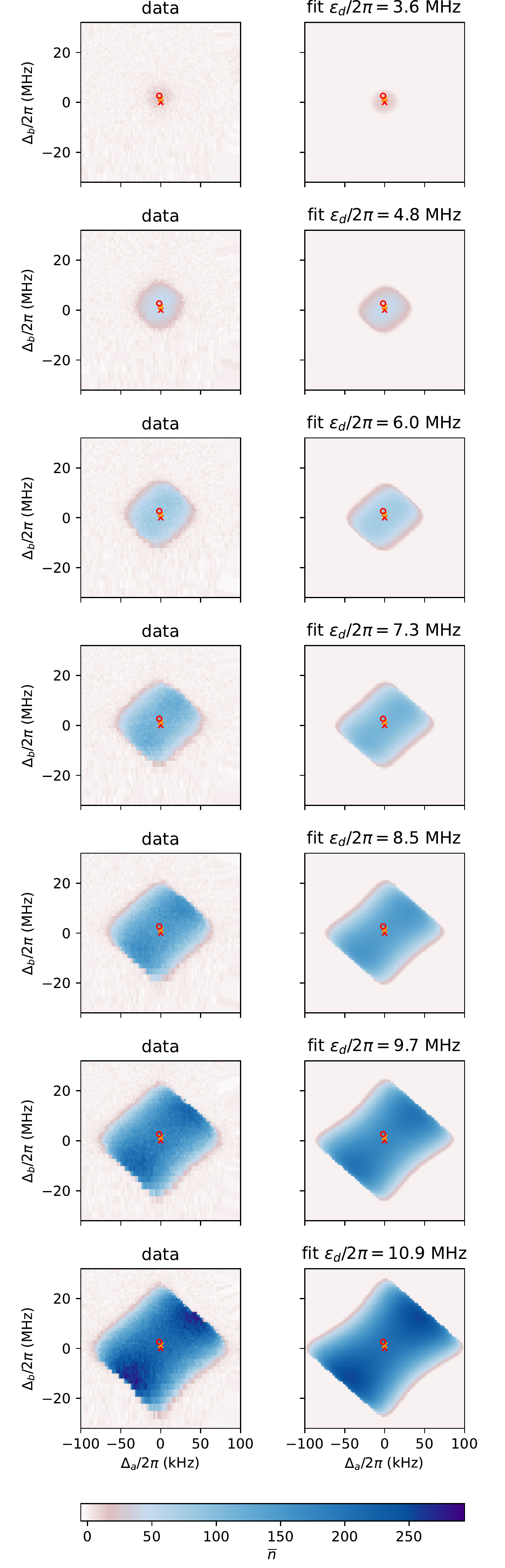}
    \caption{radiated energy from the memory in units of circulating photon number (color) as a function of the detuning from the frequency matching condition $\Delta_a$ (x-axis), and the detuning from the buffer resonance $\Delta_b$ (y-axis). Left column displays data, right column displays semi-classical simulations for the corresponding drive amplitude. Orange cross shows the position of the maximum of the auto-correlation for the largest drive amplitude. Red cross shows the zero detuning point given by direct fit of memory and buffer spectroscopy. Red circle is the point at which well averaged data were taken to perform the fit of $g_2$.}
    \label{fig:growing_diamonds}
\end{figure}

By construction, diamond center should coincide with the zero detuning point. By exploiting the diamonds inversion symmetry, we can verify that the maximum of the auto-correlation function 
\begin{equation}
\begin{split}
    &(\Delta_a^0, \Delta_b^0) \,= \\ & \max_{(\Delta_a, \Delta_b)}\left( \iint |\alpha|^2(\delta_a, \delta_b)|\alpha|^2(\Delta_a-\delta_a, \Delta_b-\delta_b) \delta_a\delta_b \right)
\end{split}
\end{equation}
gives back the zero detuning point, up to a slight discrepancy due to the diamond imperfection. Experimentally, we find the zero detuning point as the convergence point of the diamond feature at vanishingly small drive amplitude (see Fig.~\ref{fig:largediamond}c). All three centers (construction, auto-correlation, experimental) are shown in Fig.~\ref{fig:growing_diamonds} and lie in a small region at the center of the diamond.



\begin{figure}
\includegraphics[width=1\columnwidth]{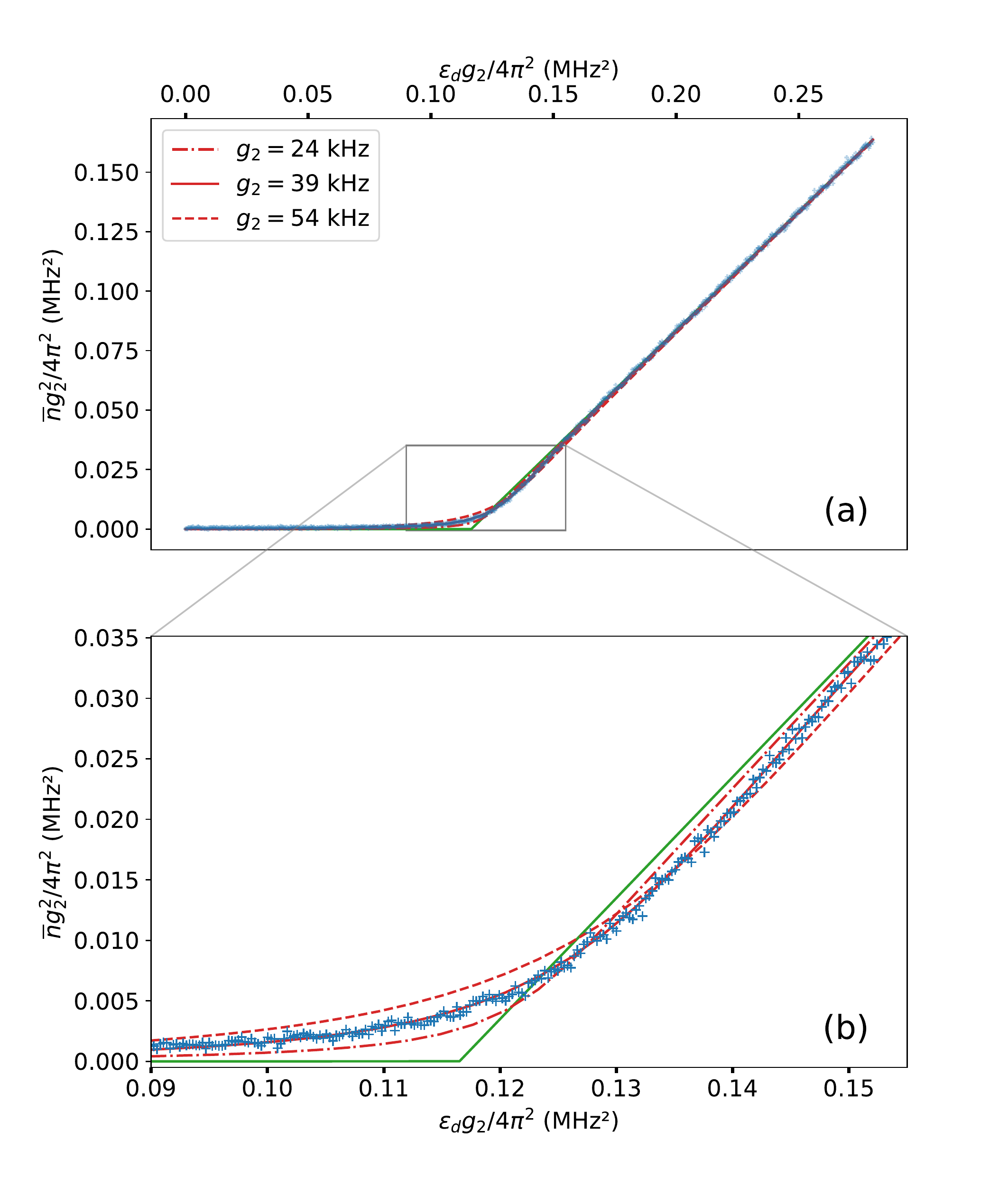}
\caption{Two photon coupling calibration. (a) radiated energy from the memory (y-axis) as a function of drive amplitude (x-axis). There are two regimes: when the drive amplitude is small, single photon loss overcome the two photon drive, and the memory stays in the vacuum. Passed the critical point, the memory gets populated by a coherent state with photon number asymptotically proportional to the drive amplitude. The axes units are chosen so that the critical point is at $\kappa_a\kappa_b/8$ and the asymptotic slope is $1$. The data correspond to an integration time of $10~\mu s$ with 10000 averages (crosses). The semi-classical model (green solid line) captures the position of the critical point but fails to explain the curvature of the experimental data. A full numerical simulation is used to reproduce the data where the only fitting parameter is $g_2$ (red lines). (b) Zoom on the curvature around the critical point (gray rectangle from (a) ), emphasizing the agreement between simulations and experimental data.}
\label{fig:fit_g_2}
\end{figure}

\section{Photon number calibration}
\label{section:supmat_photon_number_calibration}
It is of central importance that our macroscopic bit-flip times were observed for states containing only a few tens of photons. Indeed it is only in the low photon number regime that this system can operate as a coherent qubit. A reliable calibration of the number of photons $\bar{n}$ is therefore key to this work. In this section, we describe two calibration methods, and check their consistency. We start by computing the mapping between the cavity field properties and the measured quadratures. Then, we detail the method used in the main text to calibrate $\bar{n}$ from the curvature at the critical point. Finally, we describe a method relying on the measurement of the detection efficiency $\eta$.
\subsection{Heterodyne detection}
The heterodyne detection of the field radiated by the memory results in two signals that are integrated over a integration time $T_m$ to give out $(I, Q)$ pairs time traces
\begin{equation}
\label{eq:IQpairs}
\begin{split}
I_t& = \sqrt{G} \int_t^{t+T_m}\left(\sqrt{2\kappa_a^c \eta}{\rm Tr}\left(\rho_{t'} (a + a^\dag)/2\right){\rm d}t'+{\rm d}W_I\right)\\
Q_t& = \sqrt{G} \int_t^{t+T_m}\left(\sqrt{2\kappa_a^c \eta}{\rm Tr}\left(\rho_{t'} (a - a^\dag)/2i\right){\rm d}t'+{\rm d}W_Q\right)\,, 
\end{split}
\end{equation}
where $G$ is the gain of the amplification chain, $\kappa_a^c$ is the coupling rate of the memory, $\eta$ is the quantum detection efficiency, $\rho_t$ is the instantaneous state and ${\rm d} W_I$, ${\rm d} W_Q$ are the noises added to each quadrature that verify
${\rm d}W_I^2 = {\rm d}W_Q^2 = {\rm d}t$.
The statistics of the distribution of the $(I, Q)$ pairs collected over time gives information about the memory state. In particular, we can verify that in the general case \cite{Barchielli2009, Tilloy2018} and in the limit of small $T_m$
\begin{equation}
\begin{split}
    \overline{I^2 +Q^2} &= 2G T_m +  2G\kappa_a^c\eta T_m^2 \text{Tr}(\rho_\infty a^\dag a) \\
     &= 2G T_m +  2G\kappa_a^c\eta T_m^2 \bar n \\
\end{split}
\label{eq:averageI2Q2}
\end{equation}
where $\bar n$ is the mean photon number, $\overline{I^2 +Q^2}$ is the statistical average over the $(I, Q)$ pairs collected over time and $\rho_\infty$ is the steady-state density operator of the cavity. In our specific case, we have verified both numerically and experimentally that this limit is practically reached for $T_m=10$ ${\rm \mu}$s. 
In eq.~\eqref{eq:averageI2Q2}, the offset $G T_m$ can be calibrated out from the average of $\overline{I^2 +Q^2}$ when the cavity is in vacuum which results in the average energy radiated by the cavity over a period $T_m$ of
\begin{equation}
    \overline{I^2 +Q^2} - \left.\overline{I^2 +Q^2}\right|_\text{vac} = 2G\kappa_a^c\eta T_m^2 \bar n\,.
\end{equation}

\subsection{Critical point}
For various values of buffer drive amplitude $\epsilon_d$, we measure the average energy radiated by the memory for a duration $T_m=10$ ${\rm \mu}$s according to eq.~\eqref{eq:averageI2Q2} (see Fig.~\ref{fig:fit_g_2}) which is proportional to $\bar n$. The following paragraphs aim at calibrating this proportionality constant. First, we calibrate the axes such that the only unknown parameter is $g_2$ using the semi-classical analysis. Then, we use the quantum fluctuation at the critical point to determine $g_2$. Finally, we explain how we use this calibration to estimate the memory photon number in a given trajectory. 
\paragraph{Semi-classical}
As shown by equation \eqref{eq:nbar} in the semi-classical approximation, at zero detuning, the critical point appears when $|\epsilon_d g_2| = \kappa_a\kappa_b/8$ and from there the mean photon number in the memory increases linearly with the drive amplitude. Using these properties, we calibrate the drive amplitude axis in units of $|\epsilon_d g_2|$: the x-intercept of the linear dependence at large photon number (semi-classical regime) is located at $\kappa_a\kappa_b/8$. The x-axis being calibrated, we linearly stretch the y-axis such that the asymptotic slope is 1 in the strong drive regime. According to eq.~\eqref{eq:nbarg22}, this transformation enforces the y-axis to $|\alpha g_2|^2$ and leads to the scaled data of Fig.~\ref{fig:fit_g_2}.

This rescaling crucially depends on the values of $\kappa_a$ and $\kappa_b$ which are determined as follows. We measure the reflection coefficient of the memory in the presence of a pump tone slightly detuned from the frequency matching condition. This enables to capture the shift of parameters (frequency, internal losses, coupling losses) due to nonlinear effects arising from the pump while disabling the two-photon losses. From this measurement we numerically fit $\kappa_a^i/2\pi \in [15, 22]$ kHz and  $\kappa_a^c/2\pi \in [39, 42]$ kHz. The same protocol fails to determine precisely $\kappa_b$ due to the background induced by the band-stop filters and the strong dependence of the buffer parameters on the pump frequency. Instead, we use the diamond property derived in eq.~\eqref{eq:diamond_edge} that the top-right and bottom-left edge of the diamond have a slope of $-\kappa_b/\kappa_a$. We find $\kappa_b/2\pi$ in the range $[13, 20]$ MHz.

\begin{figure}
    \centering
    \includegraphics[width=0.8\columnwidth]{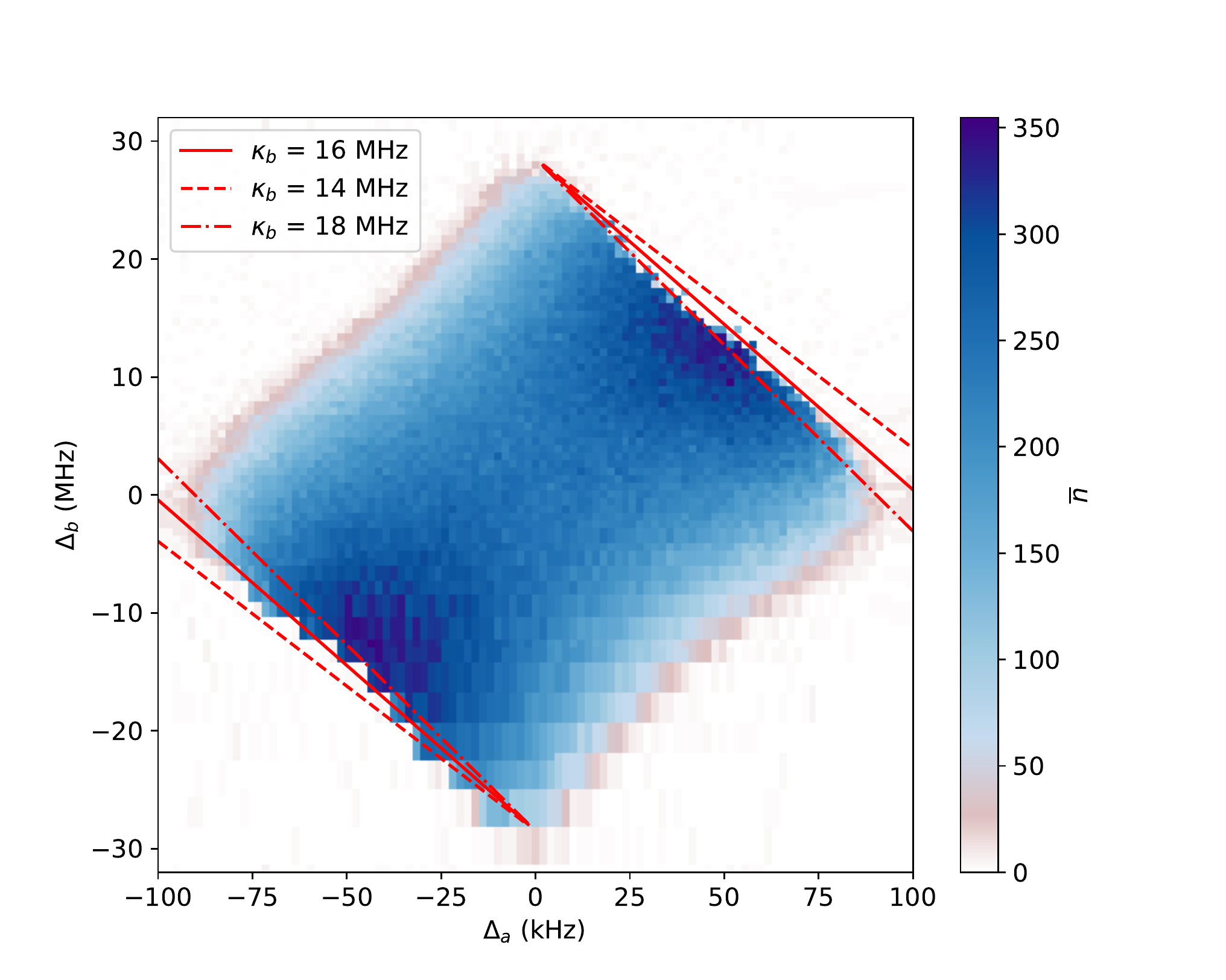}
    \caption{radiated energy from the memory in units of circulating photon number (color) as a function of the detunings $\Delta_a$ (x-axis), and $\Delta_b$ (y-axis) defined in the text, for the drive amplitude $\epsilon_d/2\pi = 12.1$ MHz. The red solid line displays the fitted slope of the top-right and bottom-left edge, yielding the ratio $\kappa_b/\kappa_a$ (see \eqref{eq:diamond_edge}). The red dashed-dotted lines and red dashed lines respectively give the upper and lower bound on this parameter (determined by graphical reading).}
    \label{fig:fit_kb_diamond}
\end{figure}

We later propagate the parameter range found for $\kappa_a$ and $\kappa_b$ on the rest of the calibration to give a robust confidence interval for $g_2$ and $\bar n$. 

We can independently check the calibration of $|\alpha g_2|^2$ by studying the excess internal losses arising from the two-photon dissipation. When the two-photon dissipation becomes resonant, the internal losses of the memory measured by direct spectroscopy increase drastically and become non-linear as a function of the probe power. The effective internal losses of the memory write 
\begin{equation}
    \kappa_{a, \text{eff}}^i = \kappa_{a}^i + 2\kappa_2|\alpha|^2
\end{equation}
where $\kappa_a^i$ is the bare internal losses of the cavity and $|\alpha|^2$ is the average circulating photon number due to the spectroscopy tone. The excess losses rewrite $8|\alpha g_2|^2/\kappa_b$ and provides an independent calibration of $|\alpha g_2|^2$ that we find in good agreement with the previous method. 

\paragraph{Quantum signature}
At the critical point, the semi-classical analysis fails to capture the curvature of the mean photon number $\bar n$ as a function of the drive amplitude $\epsilon_d$ (Fig.~\ref{fig:Fig2}). This curvature results from the quantum fluctuations at the dissipative phase transition \cite{Mylnikov2021}. Instead, we perform a quantum analysis and compute the average photon number in the steady state $\rho_\infty$ of the Lindblad equation generated by \eqref{eq:adiab_elim}, $\bar n = \text{Tr}(\rho_\infty a^\dag a)$ using the \texttt{steadystate} function imported from the QuTiP python package. Once we express $|\alpha g_2|^2$ as a function of $|\epsilon_d g_2|$ (see Fig.~\ref{fig:fit_g_2}) the only fitting parameter is $g_2$. Given the range of $\kappa_a$ and $\kappa_b$, we estimate $g_2/2\pi \in [30, 46]$ kHz.

\paragraph{Trajectory calibration}
We analyse the bit-flip time scale over several orders of magnitude, hence we increase the integration time $T_m$ to keep manageable amount of data in long bit-flip traces. Thanks to the previous calibration, we can readily get the memory photon number from the $(I, Q)$ statistics of the trace. Indeed, from eq.~\ref{eq:averageI2Q2}, we have both $G$ from the value of $\left.\overline{I^2 +Q^2}\right|_\text{vac}$ and $2G\kappa_a^c\eta$ from the calibration of $\bar n$. When given a trace with different integration time $T_m'$, we determine
\begin{equation}
\bar n = \frac{\overline{I^2 +Q^2} - 2G T_m'}{2G\kappa_a^c\eta T_m'^2}\,.
\end{equation}

\subsection{Quantum detection efficiency}
\label{sec:detection_efficiency}
A different route leading to $\bar{n}$ is to measure the detection efficiency $\eta$. Indeed, for a coherent state containing a number of photons $\bar{n}$, the measured mean $\overline{I}$ and standard deviation $\sigma(I)$ of the $I$-quadrature verify
\begin{equation}
\label{eq:eta}
\bar{n} = \left(\frac{\overline{I}}{\sigma(I)}\right)^2\frac{1}{2\eta\kappa_a^c T_m}\;.
\end{equation}
Inversely, from the calibration of $\bar{n}$ from the previous section, we estimate $\eta\simeq 3\%$.

In this section, we independently evaluate $\eta$ by fabricating a device containing a memory mode coupled to a transmon that serves as an in-situ measurement of $\bar{n}$.

\begin{figure}
    \centering
    \includegraphics[width=\columnwidth]{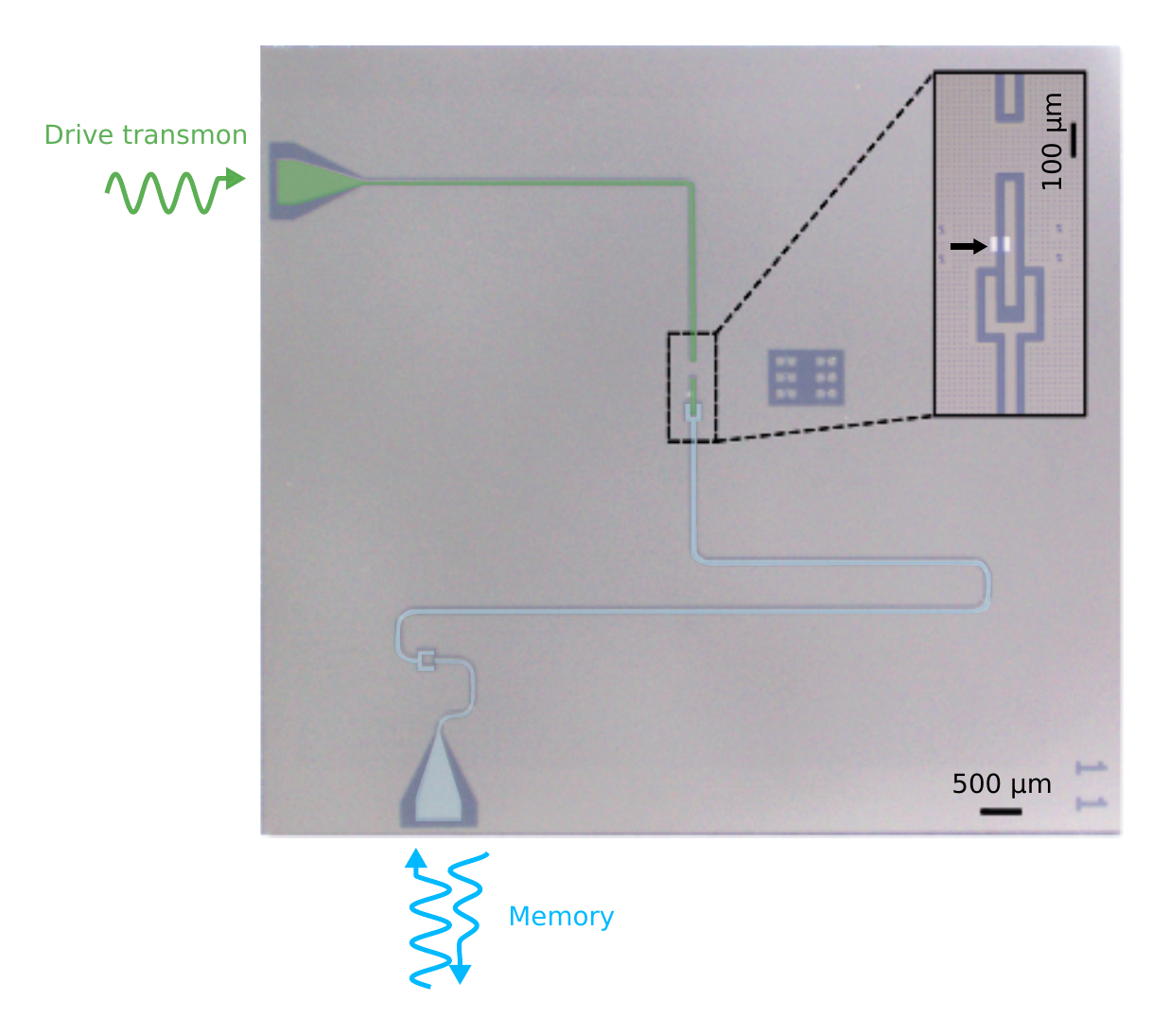}
    \caption{False-color optical micrograph of the detection efficiency chip in Nb (grey) on Si (dark blue). The memory resonator (blue) is capacitively coupled to a transmon (green).The inset is centered on the transmon and its Josephson junction in Al (light grey). We can address the memory and collect the reflected signal (blue waves) via the bottom $50\Omega$ port. The left $50\Omega$ port is dedicated to drive the transmon (green waves). This sample was also used to measure the memory thermal population of about $1\%$.}
    \label{fig:supmat_detection_efficiency}
\end{figure}

The fabricated memory mode is identical to the one described in the main text, and the entire two-photon exchange apparatus is replaced with a transmon qubit (see Fig.~\ref{fig:supmat_detection_efficiency}). The chip was mounted in a similar sample holder and measured with an identical wiring as the experiment described in the main text. The characteristics of the chip are listed in Table.~\ref{table:supmat_detection_efficiency_parameters}.

\begin{center}
\begin{table}[h]
\begin{tabular}{ |c|c| } 
 \hline
$T_1$ & $19.3 \, {\rm \mu s}$\\ 
\hline
$T_2$ & $24.3 \, {\rm \mu s}$ \\ 
 \hline
$\kappa_a^c/2\pi$ & $38\,{\rm kHz}$ \\
 \hline
 $\kappa_a^i/2\pi$ & $17\,{\rm kHz}$ \\ 
 \hline
 $\chi/2\pi$ & $1.75\,{\rm MHz}$\\ 
 \hline
\end{tabular}
\caption{Parameters of the device used to calibrate the quantum detection efficiency $\eta$. The transmon qubit lifetime and coherence times are denoted $T_1$ and $T_2$. The memory coupling and internal loss rates are denoted $\kappa_a^c$ and $\kappa_a^i$, and $\chi$ corresponds to the dispersive coupling rate between the transmon and the memory. \label{table:supmat_detection_efficiency_parameters}}
\end{table}
\end{center}
Our evaluation of $\eta$ follows three steps. First, we perform a standard spectroscopy in reflection of the memory mode in order to emulate a measurement signal that is directly proportional to the intra-cavity field amplitude $\braket{a}$. Second, for a given amplitude $a_{\rm in}$, we calibrate the cavity photon number $\bar{n}=\braket{a^\dagger a}$ by resolving the photon number splitting of the qubit. Third, for each calibrated photon number we measure the fluctuations of the outgoing field $a_{\rm out}$ and retrieve $\eta$ by inverting Eq.~\eqref{eq:eta}. We detail each step of this procedure below.

\paragraph{Memory spectroscopy} For various incoming signal amplitudes $S_{\rm in}$, we perform a spectroscopy measurement recording the reflected signal $S_{\rm out}$. Using the results of the resonance fit, we can then translate the data in the $(I,Q)$ plane in order to emulate a transmission signal: $S_{\rm t}=A\braket{a}$, where $A$ is an unknown scaling factor to be calibrated.

\begin{figure}
    \centering
    \includegraphics[width=0.9\columnwidth]{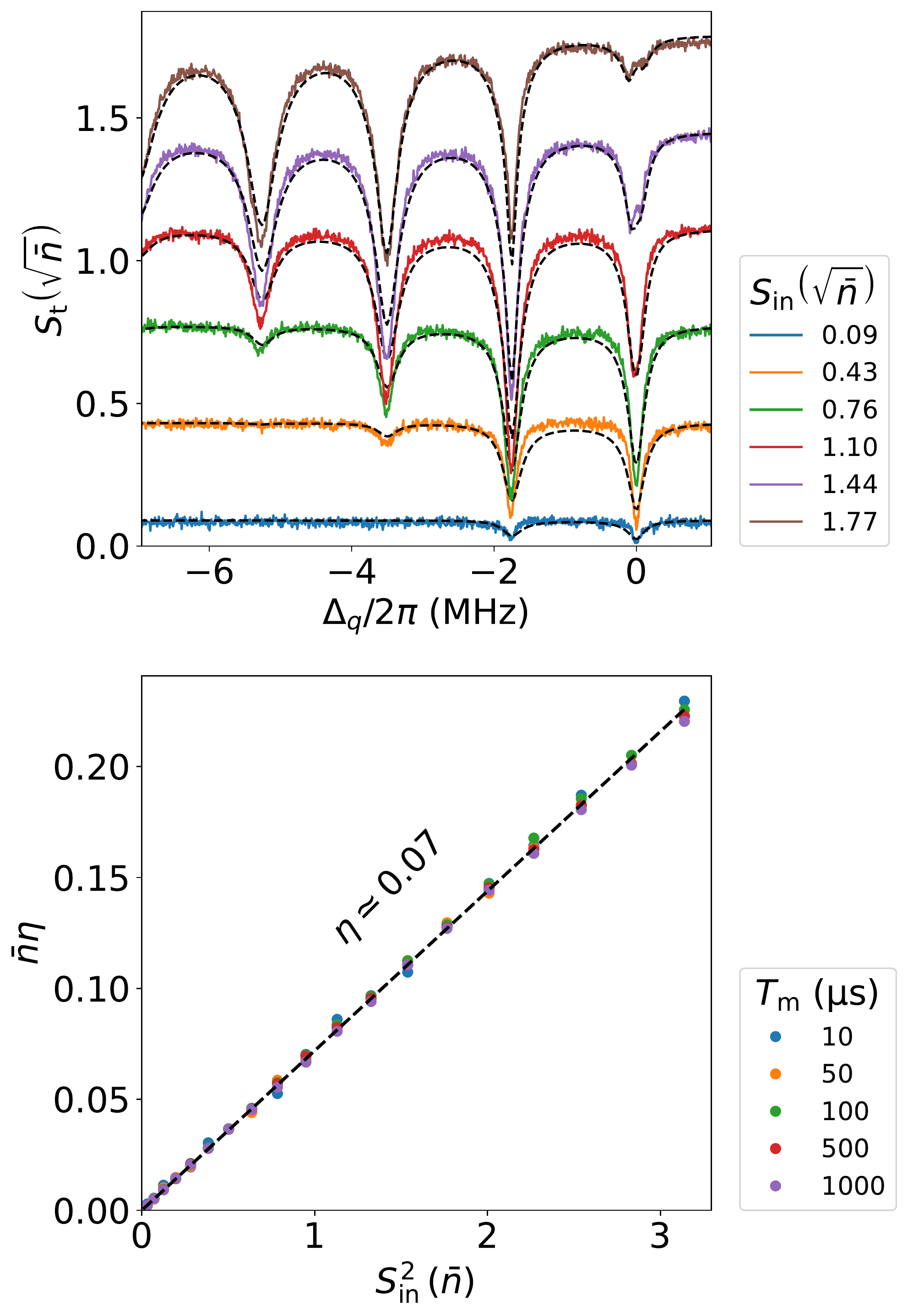}
    \caption{Calibration of the detection efficiency $\eta$. (Top) Qubit spectroscopy showing photon number splitting: data (solid lines) and fit (dashed lines).
    (Bottom) Product $\bar{n}\eta$ computed using equation \ref{eq:eta} as a function of the square input signal $S_\text{in}^2$ in units of photon number.}
    \label{fig:supmat_detection_efficiency_nbarcal}
\end{figure}

\paragraph{Photon number resolved qubit spectroscopy}
For various resonant signal amplitudes $S_{\rm in}$ we activate a drive on the transmon at a fixed amplitude $S_{q}$ with a varying detuning $\Delta_q$. The data $S_{\rm t}(\Delta_q, S_{\rm in}, S_{q})$ are then fitted to the result of a numerical simulation that we detail in the following. Using the \texttt{steadystate} function of the QuTiP package \cite{qutip1, qutip2}, we solve the following dynamics 
\begin{eqnarray*}
\partial_t{\rho} &=& -i\left[H,\rho\right] \\
&+& D\left[\sqrt{\kappa_a} a \right]\rho + D\left[\sqrt{\kappa_1} q \right]\rho + D\left[\sqrt{\kappa_{\rm \phi}} q^\dagger q \right]\rho\\
H &=& \Delta_q q^{\dagger} q - \chi  a^{\dagger}a q^{\dagger} q + \Omega_a \left(a+a^\dagger\right) + \Omega_q \left(q+q^\dagger\right)\;,
\end{eqnarray*}
where $a$ (resp. $q$) is the memory (resp. qubit) mode annihilation operator, $\kappa_1=\frac{1}{T_1}$, $\kappa_{\phi}=\frac{1}{T_2}-\frac{1}{2 T_1}$, $\Omega_a$ (resp. $\Omega_q$) is the drive on the memory (resp. qubit) and the remaining parameters are defined in table \ref{table:supmat_detection_efficiency_parameters}. From this simulation we extract $\braket{a}(\Delta_q, \Omega_a, \Omega_q)$, that is used to fit the dataset $S_{\rm t}(\Delta_q, S_{\rm in}, S_{q})$, where the fit parameters are the proportionality constants relating $S_{\rm in}$ to $\Omega_a$, $S_{q}$ to $\Omega_q$ and $S_{\rm t}$ to $\braket{a}$ (see Fig.~\ref{fig:supmat_detection_efficiency_nbarcal}).

\paragraph{Output field statistics}
For every drive amplitude $S_{\rm in}$, the previous fit estimates the intra-cavity field $\braket{a}$, and hence $\bar{n}$. By acquiring histograms of the output field $S_{\rm t}$, we now invert Eq.~\eqref{eq:eta} and retrieve $\eta\simeq 7\%$, a factor two larger than the previously estimated value. This deviation can be attributed to differences in the RF connections of the two samples. These values may be explained by lossy elements (two circulators, one Eccosorb filter and two directional couplers) between the sample and the TWPA. 

\section{Bit-flip time simulations}

\label{section:supmat_bitflip_time}
\begin{figure}[h!]
    \centering
    \includegraphics[width=0.9\columnwidth]{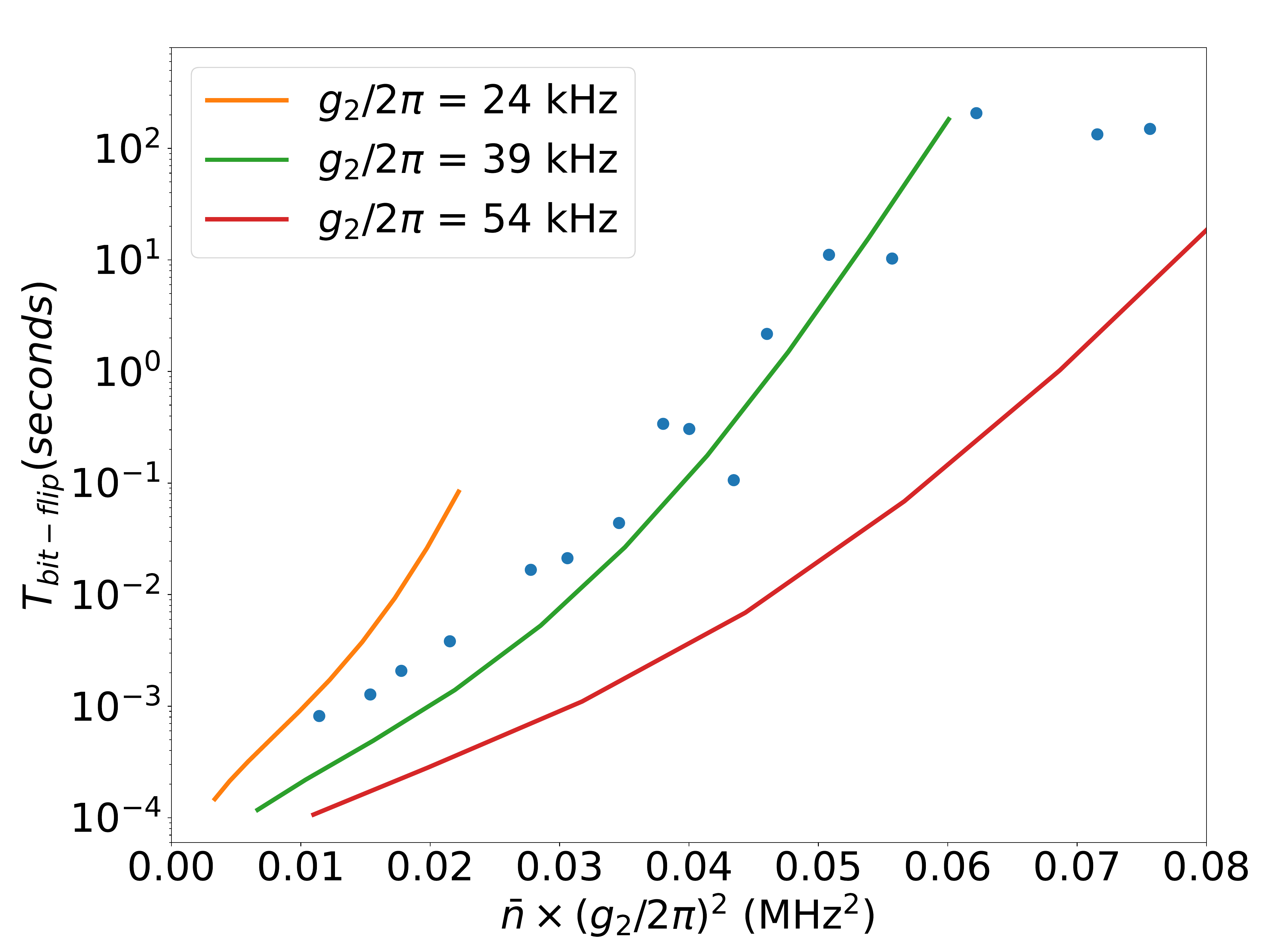}
    \caption{Numerical simulations (solid lines) of the bit-flip time (y-axis) for three values of $g_2$ (labels) as a function of the number of photons in the memory $\bar{n}$ multiplied by $(g_2/2\pi)^2$ (x-axis). The data (dots) from Fig.~\ref{fig:Fig4} of the main text qualitatively matches the simulations for $g_2/2\pi=39$~kHz, thus confirming our calibration of $g_2$.}
    \label{fig:numerics_bitflip}
\end{figure}

We numerically simulate the dynamics of the memory described in Eq.~\eqref{eq:L2L1} using the \texttt{mesolve} function imported from the QuTiP python package \cite{qutip1, qutip2}. We run the simulation for three different values of $g_2$ (or equivalently $\kappa_2$). For each of these values, we sweep $\epsilon_2$ in order to vary $\bar{n}=|\alpha|^2$ in the range of 4 to 40 photons. We initialize the memory in the coherent state $\ket{\alpha}$, and fit the expectation value of the annihilation operator $a$ to an exponentially decaying function. The extracted decay time corresponds to the bit-flip time. In Fig.~\ref{fig:numerics_bitflip}, we display the computed bit-flip time as a function of the product $\bar{n}\times (g_2/2\pi)^2$, since it is a well calibrated quantity in our experiment. The data lie in the vicinity of the simulation results for $g_2/2\pi=39$~kHz, thus confirming our calibration of $g_2$.

\end{document}